\documentclass[11pt]{article}
\usepackage{latexsym,amssymb,amsmath}
\newcommand{\ft}[2]{{\textstyle\frac{#1}{#2}}}

\def\bfone{\relax{\rm 1\kern-.35em 1}}

\newcommand{\llceil}{{|\!\!|\!\!\lceil}}
\newcommand{\rrfloor}{{\rfloor\!\!|\!\!|}}
\long\def\symbolfootnote[#1]#2{\begingroup%
\def\thefootnote{\fnsymbol{footnote}}\footnote[#1]{#2}\endgroup}

\newdimen\squaresize \squaresize=12pt
\newdimen\thickness \thickness=0.7pt

\def\square#1{\hbox{\vrule width \thickness
   \vbox to \squaresize{\hrule height \thickness\vss
      \hbox to \squaresize{\hss#1\hss}
   \vss\hrule height\thickness}
\unskip\vrule width \thickness} \kern-\thickness}

\def\cut#1{\hbox{\vrule width-1 \thickness
   \vbox to \squaresize{\hrule height-1 \thickness\vss
      \hbox to \squaresize{\hss#1\hss}
   \vss\hrule height-1\thickness}
\unskip\vrule width +4 \thickness} \kern-\thickness}

\def\vsquare#1{\vbox{\square{$#1$}}\kern-\thickness}

\begin{document}
\begin{titlepage}

\begin{flushright}\small
ITP-UU-08/73\\ SPIN-08/56
\end{flushright}
%%%%%%%%%%%%%%%%%%%%%%%%%%%%%%%%%%%%%%%%%%%%%%%%%%%%%%%%%%%%%%%
%%%%%%%%%%%%%%%%%%%%%%%%%%%%%%%%%%%%%%%%%%%%%%%%%%%%%%%%%%%%%%%
%
\vskip 10mm
\begin{center}
  {\huge \textbf{Supergravity and M-Theory}\symbolfootnote[2]{Based on a
  talk presented at {\it Quantum Gravity: Challenges and
  Perspectives}, Heraeus Seminar,\\Bad Honnef, 14-16 April, 2008.}} \\
%\end{center}
\vskip 8mm

%\begin{center}
{\large\textbf {Bernard de Wit and Maaike van Zalk}}\\[6mm]

Institute for Theoretical Physics \,\&\, Spinoza Institute,\\
Utrecht University, P.O. Box  80.195, NL-3508 TD Utrecht,
The Netherlands\\[1mm]

{{\tt b.dewit@uu.nl}\;,\;{\tt m.vanzalk@uu.nl}}
\end{center}

\vskip .2in

\begin{center} {\bf Abstract }
\end{center}
%%%%%%%%%
\begin{quotation}\noindent
  Supergravity provides the effective field theories for string
  compactifications. The deformation of the maximal supergravities by
  non-abelian gauge interactions is only possible for a restricted
  class of charges.  Generically these `gaugings' involve a hierarchy
  of $p$-form fields which belong to specific representations of the
  duality group. The group-theoretical structure of this $p$-form
  hierarchy exhibits many interesting features. In the case of maximal
  supergravity the class of allowed deformations has intriguing
  connections with M/string theory.
\end{quotation}
%%%%%%%%
%\flushleft{\today}
%%%%%%%
\end{titlepage}
\eject
%%%%%%%%%%%%%%%%%%%%%%%%%%%%%%%%%%%%%%%%%%%%%%%%%%%%%%%%%%%%
%%%%%%%%%%%%%%%%%%%%%%%%%%%%%%%%%%%%%%%%%%%%%%%%%%%%%%%%%%%%
\section{Introduction}
\setcounter{equation}{0}
\label{sec:introduction}
%%%%%%%%%%%%%%%%%%%%%%%%%%%%%%%%%%%%%%%%%%%%%%%%%%%%%%%%%%%%

Supergravity provides the effective field theories associated with
string compactifications and serves as a framework for studying a
large variety of phenomena. Among those are topics that have their
roots in general relativity, such as black holes and cosmology.
Irrespective of the precise context, supergravity itself leads to many
surprises, which, in hindsight, often have an explanation in
underlying theories, such as M-theory (an extension of string theory).
Obviously this connection is at least partly based on the presence of
non-trivial symmetries that are shared by these theories.

Here we discuss the deformations of (maximal) supergravities by
non-abelian gauge interactions and exhibit some of their connections
to M-theory. As it turns out, these deformations are quite restricted
and can be classified by group-theoretical methods. They involve a
hierarchy of $p$-form tensor fields, whose representations under the
supergravity duality group have also been obtained from M-theory in
various incarnations. The study of general gaugings of maximal
supergravities, which was initiated in
\cite{deWit:2004nw,deWit:2005hv}, led to considerable insight in the
general question of embedding a non-abelian gauge group into the rigid
symmetry group $\mathrm{G}$ of a theory that contains abelian vector
fields without corresponding charges, transforming in some
representation of $\mathrm{G}$ (usually not in the adjoint
representation). The field content of this theory is fixed up to
possible (Hodge) dualities between \mbox{$p$-forms} and
$(d-p-2)$-forms, so that it is advantageous to adopt a framework in
which the decomposition of the form fields is left open until after
specifying the gauging.

The relevance of this approach can, for instance, be seen in four
space-time dimensions \cite{deWit:2005ub}, where the Lagrangian can be
changed by electric/magnetic duality so that electric gauge fields are
replaced by their magnetic duals. In the usual setting, one has to
adopt an electric/magnetic duality frame where the gauge fields
associated with the desired gauging are all electric. In principle
this may not suffice, as the gauge fields should also decompose under
the embedded gauge group into fields transforming in the adjoint
representation of the gauge group, and fields that are invariant under
this group, so as to avoid inconsistencies.  In a more covariant
framework, on the other hand, one introduces both electric and
magnetic gauge fields from the start, such that the desired gauge
group can be embedded irrespectively of the particular
electric/magnetic duality frame. Gauge charges can then be switched on
in a fully covariant setting. Among other things this involves
introducing 2-form fields transforming in the adjoint representation
of $\mathrm{G}$. The gauge transformations associated with the 2-form
gauge fields ensure that the number of physical degrees of freedom is
not changed.

In this covariant approach the gauge group embedding is encoded in the
so-called embedding tensor, which is treated as a spurionic quantity
so as to make it amenable to group-theoretical methods. This embedding
tensor was first introduced in the context of gaugings of
three-dimensional maximal supergravity
\cite{Nicolai:2000sc,Nicolai:2001sv}.  While every choice of embedding
tensor defines a particular gauging and thereby a corresponding
$p$-form hierarchy, scanning through all possible choices of the
embedding tensor subject to certain group-theoretical representation
constraints that it must obey, enables one to characterize the
multiplicity of the various $p$-forms in entire
$\mathrm{G}$-representations -- within which every specific gauging
selects its proper subset. This is precisely the meaning of treating
the embedding tensor as a spurionic quantity.

In four space-time dimensions no $p$-form fields are required in the
action beyond ${p=2}$, but the higher-dimensional case naturally
incorporates higher-rank form fields when switching on gauge charges,
thus extending naturally to a hierarchy with a non-trivial
entanglement of forms of different ranks. It may seem that one
introduces an infinite number of degrees of freedom in this way, but,
as mentioned already above, the hierarchy contains additional gauge
invariances beyond those associated with the vector fields.  This
$p$-form hierarchy is entirely determined by the rigid symmetry group
$\mathrm{G}$ and the embedding tensor that defines the gauge group
embedding into $\mathrm{G}$ \cite{deWit:2005hv,deWit:2008ta} and a
priori makes no reference to an action nor to the number $d$ of
space-time dimensions.  As a group-theoretical construct, the $p$-form
hierarchy continues indefinitely, but in practice it can be
consistently truncated in agreement with the space-time properties
(notably the absence of forms of a rank $p>d$).

In the context of a given Lagrangian the details of the $p$-form
hierarchy will change and the transformation rules are deformed by the
presence of various matter fields. As a result, the closure of the
generalized gauge algebra may involve additional symmetries. The
hierarchy may turn out to be truncated at a relatively early stage,
because the Lagrangian may be such that the gauge transformations that
connect to the higher-$p$ forms have become trivially satisfied.  On
the other hand, the $(d\!-\!1)$- and $d$-forms play a different role,
as was suggested in \cite{deWit:2008ta}, where this was explicitly
demonstrated for three-dimensional maximal supergravity.

In this paper we review a number of elements of the $p$-form hierarchy
and its connections to M-theory. In section
\ref{sec:kaluza-klein-theory} we give a qualitative introduction to
the so-called hidden symmetries that emerge in torus compactifications
of higher-dimensional gravity theories, to appreciate some of the
duality symmetries that are relevant for the maximal supergravities.
In a separate subsection \ref{sec:gauge-deformations}, we demonstrate
how the $p$-form fields appear upon the introduction of non-abelian
gauge interactions.  Section \ref{sec:p-form-assignments} first
describes the pattern obtained for the $p$-form representations for
maximal supergravity with space-time dimensions $d=3,\ldots,7$. In a
separate subsection \ref{sec:more-about-hierarchy} we then try to
generalize this pattern and show that it is in fact more generic.  In
a second subsection \ref{sec:m-theory} we describe how the
representation content of the $p$-form gauge fields can be connected
to results obtained in M-theory in a completely different context. In
section \ref{sec:hierarchyin4d} we deal with generic gauge theories in
four space-time dimensions in order to explain a number of features
relevant for the $p$-form hierarchy in some more detail.

%%%%%%%%%%%%%%%%%%%%%%%%%%%%%%%%%%%%%%%%%%%%%%%%%%%%%%%%%%
\section{Kaluza-Klein theory and gauge deformations}
\label{sec:kaluza-klein-theory}
\setcounter{equation}{0}
%%%%%%%%%%%%%%%%%%%%%%%%%%%%%%%%%%%%%%%%%%%%%%%%%%%%%%%%%%
Supergravity is an extension of general relativity that, in addition,
is invariant under local supersymmetry, which transforms fermionic
into bosonic fields and vice versa.  Assuming that supergravity is an
interacting field theory based on a finite number of fields, and that
it allows a flat Minkowski space-time with maximal supersymmetry as a
solution, it can be realized in at most $D=11$ space-time dimensions
with the number of independent supersymmetries restricted to 32.
Supergravity in eleven space-time dimensions \cite{Cremmer:1978km}
involves only three fields, namely a graviton field $g_{\mu\nu}$, a
3-form gauge field $A_{\mu\nu\rho}$ and a gravitino field $\psi_\mu$.
Space-time indices are consistently denoted by $\mu,\nu, \ldots$ and,
in this case, take the values $\mu,\nu,\ldots=0, 1,2,\ldots,10$. In
eleven space-time dimensions spinors carry 32 components, so that
$\psi_\mu$ is a 32-component vector spinor. Maximal supergravity
theories in $d$ space-time dimensions can be obtained by compactifying
$11-d$ dimensions on a hyper-torus $T^{11-d}$. These compactified
theories exhibit a remarkable invariance group of so-called `hidden
symmetries'. Deformations of the toroidally compactified theories are
generally possible by introducing gauge interactions whose
corresponding gauge group is embedded into the hidden symmetry group.
This will be the topic of the second subsection
\ref{sec:gauge-deformations}. In a few specific cases alternative
deformations are possible as well.  Of course, compactifications on
non-flat manifolds can also be considered, and most of them will
involve a breaking of supersymmetry. Some of these compactifications
may arise as the result of a gauge deformation of a toroidally
compactified theory, but this possibility will be ignored below.

In the first subsection \ref{sec:hidden-symmetries} we discuss the
emergence of the `hidden symmetry' group, first in gravity possibly
extended with a tensor and a scalar field, and subsequently in
supergravity.

%%%%%%%%%%%%%%%%%%%%%%%%%%%%%%%%%%%%%%%%%%%%%%%%%%%%%%%
\subsection{Hidden symmetries}
\label{sec:hidden-symmetries}
%%%%%%%%%%%%%%%%%%%%%%%%%%%%%%%%%%%%%%%%%%%%%%%%%%%%%%%
As an introduction we first discuss toroidal compactifications of
general relativity, or an extension thereof, following the approach of
Kaluza and Klein, who originally started from five space-time dimensions
\cite{KK}.  We first discuss the so-called `hidden symmetries',
demonstrating how the corresponding symmetry group takes a more
interesting form upon including additional fields in the
higher-dimensional theory.  Subsequently we restrict ourselves to the
massless sector and study general deformations of these theories by
introducing additional gauge interactions.  Hence, consider general
relativity in $D$ space-time dimensions, with $n$ dimensions
compactified on the torus $T^n$, so that space-time decomposes
according to
\begin{equation}
  \label{eq:mxtorus}
  \mathcal{M}_{D} \to \mathcal{M}_d\times T^n\,,
\end{equation}
where $d=D-n$. The resulting $d$-dimensional theory then describes
massless graviton states, $n$ abelian gauge fields (called
Kaluza-Klein photons) and $\tfrac12 n(n+1)$ massless scalar fields, as
well as an infinite tower of massive graviton states. Besides the
$d$-dimensional general coordinate transformations and the abelian
gauge transformations, the theory turns out to be invariant under the
group $\mathrm{GL}(n)$, which is non-linearly realized on the massless
scalar fields. The latter fields parameterize the
$\mathrm{GL}(n)/\mathrm{SO}(n)$ maximally symmetric space.  The
massive fields are all charged and couple to the $n$ abelian gauge
fields with quantized charges. This restricts the $\mathrm{GL}(n)$
invariance to a discrete subgroup $\mathrm{GL}(n,\mathbb{Z})$ which
leaves the lattice of Kaluza-Klein charges invariant.

The pattern of dimensional compactification changes when the dimension
$d$ of the lower-dimensional space-time becomes equal to three. In
three space-time dimensions, gravitons no longer carry local degrees
of freedom (only topological ones) and the degrees of freedom residing
in the Kaluza-Klein photons can be carried by scalar fields (here and
henceforth we suppress the massive fields to which these photons
couple, and concentrate on the massless sector). Hence the massless
sector of the theory can be entirely formulated in terms of $\tfrac12
n(n+3)$ scalar fields. The symmetry group is now extended from
$\mathrm{GL}(n)$ to $\mathrm{SL}(n+1)$ (which are of equal rank) and
the scalar fields parameterize the space
$\mathrm{SL}(n+1)/\mathrm{SO}(n+1)$, which reflects the extended
symmetry.

The emergence of hidden symmetries is a well known phenomenon in
dimensional compactification. The rank of the symmetry group in $d$
dimensions is always increased by $n$ as compared to the rank of the
symmetry group in the original $D$-dimensional theory, where $n=D-d$
denotes the number of toroidally compactified dimensions. Part of
these hidden symmetries can be derived directly from a subset of the
gauge transformations in the higher-dimensional ancestor theory, but
others are somewhat less obvious. In toroidal compactifications it can
also be shown that, when the massless scalars parameterize a
homogeneous space in higher dimensions, this will also be the case in
lower dimensions.

The presence of tensor gauge fields in the $D$-dimensional theory
introduces further structure. To demonstrate this, consider, for
instance, the Lagrangian of general relativity coupled to an
anti-symmetric 2-form tensor field $B_{\mu\nu}$ in $D$ space-time
dimensions,
\begin{equation}
  \label{eq:gravitensor-D}
  \mathcal{L}_D =  -\tfrac12\sqrt{g}\,R -\tfrac34 \sqrt{g}\,
  \left(\partial_{[\mu}B_{\nu\rho]}\right)^2 \,.
\end{equation}
Its toroidal compactification leads to the symmetry group
$\mathrm{SO}(n,n;\mathbb{Z})$, which again has rank $n$. The
lower-dimensional theory describes massless states belonging to the
graviton, the antisymmetric tensor, and $2n$ spin-1 and $n^2$ spin-0
states. The massless scalars parameterize the space
$\mathrm{SO}(n,n;\mathbb{R})/
[\mathrm{SO}(n;\mathbb{R})\times\mathrm{SO}(n;\mathbb{R})]$.
Furthermore there will be a tower of massive graviton and
antisymmetric tensor states.  This generic pattern will now change in
space-time dimensions $d\leq5$.  In $d=5$ dimensions an antisymmetric
tensor gauge field can be dualized to a vector gauge field, whereas,
in $d=4$ dimensions, a tensor gauge field can be converted into a
scalar field.

To make the theory a bit more interesting, let us also include a
scalar field in $D$ dimensions, which couples such that the theory
is invariant under certain scale transformations. This means that the
theory in $D$ dimensions has an invariance group of unit rank.
Depending on how precisely this scalar field interacts, the following
result may arise in $d$ dimensions (always assuming $n=D-d)$,
\begin{equation}
  \label{eq:vector-tensor-toro-n}
  \begin{array}{rcll}
    d>5& : & \mathrm{G}=\mathbb{R}^+\times \mathrm{SO}(n,n;\mathbb{Z})
    & (n,n)\;
    \mathrm{vectors}\\
    d=5 & : &\mathrm{G}=\mathbb{R}^+\times\mathrm{SO}(n,n;\mathbb{Z})
    & (n,n)+1\; \mathrm{vectors}    \\
    d=4 & :
    &\mathrm{G}=\mathrm{SL}(2;\mathbb{Z})\times\mathrm{SO}(n,n;\mathbb{Z})
    & (n,n)+1\; \mathrm{vectors}    \\
    d=3 & : & \mathrm{G}=\mathrm{SO}(n+1,n+1;\mathbb{Z}) &
    \mathrm{no\;vectors}
  \end{array}
\end{equation}
All these symmetry groups have rank $n+1$, in agreement with the
general theorem. The massless scalars always parameterize a homogeneous
space, namely $\mathrm{SO}(n,n;\mathbb{R})/
[\mathrm{SO}(n;\mathbb{R})\times\mathrm{SO}(n;\mathbb{R})]$, which,
for $d=4$, is multiplied by a
$\mathrm{SL}(2;\mathbb{R})/\mathrm{SO}(2;\mathbb{R})$ factor.

%%%%%%%%%%%%%%%%%%%%%%%%%%%%%%%%%%%%%%%%%%%%%%%%%%%%%%%%%%%%%%%%%%
%%%%%%%%%%%%%%%%%%%%%%%%%%%%%%%%%%%%%%%%%%%%%%%%%%%%%%%%%%%%%%%%%%
\begin{table}
\begin{center}
\begin{tabular}{l l  l l  }\hline
~&~&~&~\\[-4mm]
$d$ &${\rm G}$& ${\rm H}$ & $\Theta$  \\   \hline
~&~&~&~\\[-4mm]
7   & ${\rm SL}(5)$ & ${\rm USp}(4)$  & ${\bf 10}\times {\bf 24}= {\bf
  10}+\underline{\bf 15}+  \underline{\bf 40}+ {\bf 175}$  \\[1mm]
6  & ${\rm SO}(5,5)$ & ${\rm USp}(4) \times {\rm USp}(4)$ &
  ${\bf 16}\times{\bf 45} =
  {\bf 16}+ \underline{\bf 144} + {\bf 560}$ \\[.8mm]
5   & ${\rm E}_{6(6)}$ & ${\rm USp}(8)$ & ${\bf 27}\times{\bf 78} =
  {\bf 27} + \underline{\bf 351} + {\bf 1728}$  \\[.5mm]
4   & ${\rm E}_{7(7)}$ & ${\rm SU}(8)$  & ${\bf 56}\times{\bf 133} =
  {\bf 56} + \underline{\bf 912} + {\bf 6480}$   \\[.5mm]
3   & ${\rm E}_{8(8)}$ & ${\rm SO}(16)$ & ${\bf 248}\times{\bf 248} =
  \underline{\bf 1} + {\bf 248} + \underline{\bf 3875} +{\bf 27000}
  +  {\bf 30380}$
\\ \hline
\end{tabular}
\end{center}
\caption{\small
Decomposition of the embedding tensor $\Theta$ for maximal
supergravities in various space-time dimensions in terms of irreducible
${\rm G}$ representations.  Only the underlined representations are
allowed according to the representation constraint. The R-symmetry
group ${\rm H}$ is the maximal compact subgroup of ${\rm G}$.
}\label{tab:T-tensor-repr}
\end{table}
%%%%%%%%%%%%%%%%%%%%%%%%%%%%%%%%%%%%%%%%%%%%%%%%%%%%%%%%%%%%%%%%%%
%%%%%%%%%%%%%%%%%%%%%%%%%%%%%%%%%%%%%%%%%%%%%%%%%%%%%%%%%%%%%%%%%%

As is well known, supergravity leads to a large variety of such hidden
symmetry groups, including some of the exceptional groups. This is
shown in table \ref{tab:T-tensor-repr}, where we list the symmetry
group $\mathrm{G}$ and its maximal compact subgroup $\mathrm{H}$ for
maximal supergravities in space-time dimensions $d= 3,\ldots,7$. In
supergravity the symmetry group is usually called the duality group,
and in this context we will consistently use this nomenclature. The
massless scalar fields then parameterize the homogeneous space
$\mathrm{G}/\mathrm{H}$, and $\mathrm{H}$ coincides with the
so-called R-symmetry group. The latter is the subgroup of the
automorphism group of the supersymmetry algebra that commutes with the
$d$-dimensional Lorentz transformations.

What we will be interested in is to study all possible deformations of
supergravity theories that are induced by non-abelian gauge
interactions. The corresponding gauge group must obviously be a
subgroup of the duality group. The gauge fields, which so far were
abelian and coupled only to the massive modes, will now couple also to
the massless fields.  Henceforth the massive modes will be
discarded. For lower dimensions one obviously faces a dilemma, because
by dualizing the higher-rank form fields one increases the degree of
symmetry. On the other hand one also decreases the number of gauge
fields in this way, and thus seems to constrain the possible gauge
groups. The most extreme example of this is encountered in three
space-time dimensions, where all the gauge fields can in fact be
dualized to scalars, upon which the hidden symmetry group is extended,
but on the other hand, there seem to be no gauge fields left to induce
the gauging.

This last puzzle was resolved in \cite{Nicolai:2000sc,Nicolai:2001sv},
for maximal supergravity in three space-time dimensions. This theory
does not contain any vector fields, and has the symmetry group
$\mathrm{E}_{8(8)}(\mathbb{R})$.\footnote{%%%%%%%%%%%%%%%%%%%%%%
  A similar approach was followed subsequently for non-maximal
  supergravity in three space-time dimensions \cite{deWit:2003ja} (see
  also \cite{deWit:2004yr}). } %%%%%%%%%%%%%%%%%%%%%%%%%%%%%%%
Namely one adds a Chern-Simons term for as many gauge fields as there
are rigid (continuous) invariances. For maximal supergravity in three
dimensions this implies that one introduces 248 gauge fields,
associated with the hidden symmetry group
$\mathrm{E}_{8(8)}(\mathbb{R})$,
\begin{equation}
  \label{eq:CS-E8}
  \mathcal{L}_\mathrm{CS}\propto g\, \varepsilon^{\mu\nu\rho} A_\mu{}^M
  \,\Theta_{MN} \left[\partial_\nu A_\rho{}^N -\tfrac13 g\,f_{PQ}{}^N
  A_\nu{}^P A_\rho{}^Q \right]\,,
\end{equation}
where $g$ is the gauge coupling constant, $f_{PQ}{}^N$ denotes the
structure constants associated with a subgroup of
$\mathrm{E}_{8(8)}(\mathbb{R})$, and $M,N,\ldots= 1,2,\ldots,248$.
Furthermore, $\Theta_{MN}$ is a numerical symmetric tensor named the
{\it embedding tensor}, which encodes the embedding of the gauge group
in $\mathrm{E}_{8(8)}(\mathbb{R})$. The structure constants
$f_{MN}{}^P$ are implicitly encoded in the embedding tensor, in a way
that we will exhibit below. In the next subsection we will introduce
the embedding tensor in a more general context, where it is not
necessarily a symmetric tensor. As we shall see, this set-up leads
naturally to the introduction of a hierarchy of $p$-form fields.

%%%%%%%%%%%%%%%%%%%%%%%%%%%%%%%%%%%%%%%%%%%%%%%%%%%%%%%%%%%%%%%%
\subsection{Gauge deformations}
\label{sec:gauge-deformations}
%%%%%%%%%%%%%%%%%%%%%%%%%%%%%%%%%%%%%%%%%%%%%%%%%%%%%%%%%%%%%%%%
We start with a theory with abelian gauge fields
$A_\mu{}^{M}$, that is invariant under a group $\mathrm{G}$ of rigid
transformations. The gauge fields transform in a representation
$\mathcal{R}_\mathrm{v}$ of that group.\footnote{%%%%%%%%%%%%%%%%%%
  In even space-time dimensions this assignment may fail and complete
  $\mathrm{G}$ representations may require the presence of magnetic
  duals. This was first demonstrated in \cite{deWit:2005ub} in four
  space-time dimensions.} %%%%%%%%%%%%%%%%%%%%%%%%%%%%%%%%%%%%%%%%%%
The generators in this representation are denoted by
$(t_\alpha)_{M}{}^{N}$, so that $\delta A_\mu{}^
{M}=-\Lambda^\alpha(t_\alpha)_{N}{}^{M} \,A_\mu{}^{N}$, and the
structure constants $f_{\alpha\beta}{}^\gamma$ of $\mathrm{G}$ are
defined according to $[t_\alpha,t_\beta]= f_{\alpha\beta}{}^\gamma
\,t_\gamma$. The next step is to select a subgroup of $\mathrm{G}$
that will be elevated to a gauge group with non-trivial gauge charges,
whose dimension is obviously restricted by the number of vector
fields.  The discussion in this section will remain rather general and
will neither depend on $\mathrm{G}$ nor on the space-time dimension.
We refer to \cite{Nicolai:2000sc,deWit:2004nw,Samtleben:2005bp,
  deWit:2007mt,Bergshoeff:2007ef} where a number of results was
described for maximal supergravity in various dimensions.

The gauge group embedding is defined by specifying its generators $X_
{M}$,\footnote{%%%%%%%%%%%%%%%%%%%%%%%%%%%%%%%%%%%%%%%%%%%%%%%%%%%%
  The corresponding gauge algebra may have a central extension acting
  exclusively on the vector fields. }  %%%%%%%%%%%%%%%%%%%%%%%%%%%%
which couple to the gauge fields $A_\mu{}^ {M}$ in the usual fashion,
and which can be decomposed in terms of the independent
$\mathrm{G}$-generators $t_\alpha$, i.e.,
\begin{equation}
  \label{eq:X-theta-t}
  X_ {M} = \Theta_ {M}{}^\alpha\,t_\alpha \;,
\end{equation}
where $\Theta_ {M}{}^{\alpha}$ is the {\it embedding tensor}
transforming according to the product of the representation conjugate
to $\mathcal{R}_\mathrm{v}$, the representation in which the gauge
fields transform, and the adjoint representation of $\mathrm{G}$.
This product representation is reducible and decomposes into a number
of irreducible representations.  Only a subset of these
representations is allowed.  For supergravity the precise constraints
on the embedding tensor follow from supersymmetry, but from all
applications worked out so far, we know that at least part (if not
all) of the representation constraints is also required for purely
bosonic reasons, such as gauge invariance of the action and
consistency of the tensor gauge algebra.  This constraint on the
embedding tensor is known as the {\it representation constraint}. In
table~\ref{tab:T-tensor-repr} we have also included the representation
constraints for maximal supergravity with $d=3,\ldots,7$. It is
important to note that we will always treat the embedding tensor as a
spurionic object, which we allow to transform under $\mathrm{G}$, so
that the Lagrangian and transformation rules remain formally
$\mathrm{G}$-invariant. Only at the end we will freeze the embedding
tensor to a constant, so that the $\mathrm{G}$-invariance will be
broken.  As was shown in \cite{deWit:2008ta,deWit:2008gc} this last
step can also be described in terms of a new action in which the
freezing of $\Theta_ {M}{}^\alpha$ will be the result of a more
dynamical process. This will be discussed in due course.

The embedding tensor must satisfy a second constraint, the so-called
{\em closure constraint}, which is quadratic in
$\Theta_ {M}{}^\alpha$ and more generic. This constraint
ensures that the gauge transformations form a group so that the
generators (\ref{eq:X-theta-t}) will close under commutation. Any
embedding tensor that satisfies the closure constraint, together with
the representation constraint mentioned earlier, defines a consistent
gauging. The closure constraint reads as follows,
\begin{equation}
  \label{eq:gauge-inv-embedding}
  \mathcal{Q}_{PM}{}^\alpha=
   \Theta_{  P}{}^\beta t_{\beta  {M}}{}^{  N}
    \Theta_{  N}{}^\alpha +
    \Theta_{  P}{}^\beta f_{\beta\gamma}{}^\alpha
    \Theta_{  M}{}^\gamma \approx 0 \,,
\end{equation}
and can be interpreted as the condition that the embedding tensor
should be invariant under the embedded gauge group. Hence we can write
the closure constraint as,
\begin{equation}
  \label{eq:constraint2}
  \mathcal{Q}_{MN}{}^\alpha \equiv \delta_M\Theta_N{}^\alpha=
  \Theta_M{}^\beta \,\delta_\beta\Theta_N{}^\alpha \approx 0  \,,
\end{equation}
where $\delta_M$ and $\delta_\alpha$ denote the effect of an
infinitesimal gauge transformation or an infinitesimal
$\mathrm{G}$-transformation, respectively. We indicate that
$\mathcal{Q}_{MN}{}^\alpha$ is ``weakly zero''
($\mathcal{Q}_{MN}{}^\alpha\approx 0$) because later on we will
introduce a description where the closure constraint will be imposed by
certain field equations.
Contracting (\ref{eq:gauge-inv-embedding}) with $t_\alpha$ leads to,
\begin{equation}
  \label{eq:XX-commutator}
  {[X_ {M},X_ {N}]} \approx   -X_ {MN}{}^ {P}\,X_ {P} \,.
\end{equation}

It is noteworthy here that the generator $X_{MN}{}^{P}$ and the
structure constants of the gauge group are related, but do not
have to be identical. In particular $X_ {MN}{}^{P}$ is in general not
antisymmetric in $[{MN}]$. The embedding tensor acts as a projector,
and only in the projected subspace the matrix $X_ {MN}{}^ {P}$ is
antisymmetric in $[MN]$ and the Jacobi identity will be satisfied.
Therefore (\ref{eq:XX-commutator}) implies in particular that
$X_{({MN})}{}^{P}$ must vanish when contracted with the embedding
tensor. Denoting
 \begin{equation}
  \label{eq:def-Z}
  Z^{P}{}_{{MN}} \equiv
  X_{({MN})}{}^ {P} \,,
\end{equation}
this condition reads,
\begin{equation}
  \label{eq:closure}
  \Theta_P{}^\alpha \, Z^{P}{}_{MN} = \mathcal{Q}_{(MN)}{}^\alpha
  \approx 0\,.
\end{equation}
The tensor $Z^{P}{}_{{MN}}$ is constructed by contraction of the
embedding tensor with $\mathrm{G}$-invariant tensors and therefore
transforms in the same representation as $\Theta_ {M}{}^\alpha$ ---
except when the embedding tensor transforms reducibly so that
$Z^{P}{}_{MN}$ may actually depend on a smaller representation.  The
closure constraint~(\ref{eq:constraint2}) ensures that
$Z^{P}{}_{{MN}}$ is gauge invariant.  As is to be expected,
$Z^{P}{}_{{MN}}$ characterizes the lack of closure of the generators
$X_{M}$.  This can be seen, for instance, by calculating the direct
analogue of the Jacobi identity (in the remainder of this section we
assume that the closure constraint is identically satisfied),
\begin{equation}
\label{Jacobi-X}
   X_{[{NP}}{}^{R}\,X_{ {Q}]{R}}{}^{M} =
   \ft23 Z^ {M}{}_{ {R}[ {N}}\,  X_{ {PQ}]}{}^ {R} \,.
\end{equation}
The fact that the right-hand side does not vanish has direct
implications for the non-abelian field strengths: the standard
expression
\begin{equation}
  \label{eq:field-strength}
  \mathcal{F}_{\mu\nu}{}^ {M} =\partial_\mu A_\nu{}^ {M}
  -\partial_\nu
  A_\mu{}^ {M} + g\, X_{NP}{}^ {M}
  \,A_{[\mu}{}^{N} A_{\nu]}{}^ {P} \,,
\end{equation}
which appears in the commutator
$[D_\mu,D_\nu]= - g  \mathcal{F}_{\mu\nu}{}^ {M}\,X_ {M}$
of covariant derivatives
\begin{equation}
  \label{eq:vector-gauge-tr}
 D_\mu ~\equiv~ \partial_\mu -g\, A_\mu{}^ {M}
  \,X_ {M} \,,
\end{equation}
is not fully covariant. Rather, under standard gauge transformations
\begin{equation}
  \label{eq:A-var}
  \delta A_\mu{}^ {M} =  D_\mu\Lambda^ {M} =
  \partial_\mu \Lambda^ {M} + g A_\mu{}^ {N}
  X_ {NP}{}^ {M} \Lambda^ {P} \,,
\end{equation}
the field strength $\mathcal{F}_{\mu\nu}{}^M$ transforms as
\begin{eqnarray}
  \label{eq:delta-cal-F}
  \delta\mathcal{F}_{\mu\nu}{}^ {M}
  &=&2\, D_{[\mu}\delta
  A_{\nu]}{}^ {M} -
  2 g\, Z^{M}{}_{PQ} \,A_{[\mu}{}^ {P}
  \,\delta A_{\nu]}{}^ {Q} \nonumber\\[1ex]
  &=&  g\, \Lambda^ {P}
  X_ {NP}{}^ {M}
  \, \mathcal{F}_{\mu\nu}{}^ {N} - 2 g\, Z^ {M}{}_ {PQ}
  \,A_{[\mu}{}^ {P}\,\delta A_{\nu]}{}^ {Q}  \,.
\end{eqnarray}
This expression is {\em not} covariant, not only because of the
presence of the second term on the right-hand side, but also because
the lack of antisymmetry of $X_ {NP}{}^ {M}$ prevents us from
obtaining the expected result by inverting the order of indices $NP$
in the first term on the right-hand side.  As a consequence, we cannot
use $\mathcal{F}_{\mu\nu}{}^{M}$ in the Lagrangian, because one
needs suitable covariant field strengths for the invariant kinetic
term of the gauge fields.

To remedy this lack of covariance, the strategy followed
in~\cite{deWit:2004nw,deWit:2005hv} has been to introduce additional
(shift) gauge transformations on the vector fields,
\begin{equation}
  \label{eq:A-var-2}
\delta A_\mu{}^ {M} =  D_\mu\Lambda^ {M} -
g\,Z^ {M}{}_{ {NP}}\,\Xi_\mu{}^{ {NP}}\,,
\end{equation}
where the transformations proportional to $\Xi_\mu{}^ {NP}$ enable one
to gauge away those vector fields that are in the sector of the gauge
generators $X_ {MN}{}^ {P}$ in which the Jacobi identity is not
satisfied (this sector is perpendicular to the embedding tensor by
(\ref{eq:closure})).  Fully covariant field strengths can then be
defined upon introducing 2-form tensor fields $B_{\mu\nu}{}^ {NP}$
belonging to the same representation as $\Xi_\mu{}^{ {NP}}$,
\begin{equation}
  \label{eq:modified-fs}
{\cal H}_{\mu\nu}{}^{ {M}} =
{\cal F}_{\mu\nu}{}^ {M}  + g\, Z^ {M}{}_ {NP}
\,B_{\mu\nu}{}^ {NP}\;.
\end{equation}
These tensors transform covariantly under
gauge transformations
\begin{eqnarray}
\delta\,\mathcal{H}_{\mu\nu}{}^{{M}}
&=&
 -g\Lambda^{P}
  X_{ {P}{N}}{}^{M}
  \mathcal{H}_{\mu\nu}{}^ {N} \,,
\end{eqnarray}
provided we impose the following transformation law for the 2-forms
\begin{equation}
  \label{eq:cov-delta-B}
  Z^M{}_{NP}\; \delta B_{\mu\nu}{}^{{NP}}
  ~=~ Z^M{}_{NP}\,\Big(
   2\,D_{[\mu}\Xi_{\nu]}{}^{ {NP}} -2\,
  \Lambda^{ {N}}\mathcal{H}_{\mu\nu}{}^{ {P}}
   + 2\, A_{[\mu}{}^{ {N}}
  \delta A_{\nu]}{}^{{P}} \Big) \,.
\end{equation}
We note that the constraint (\ref{eq:closure}) ensures that
\begin{equation}
  \label{eq:ricci}
{} [D_\mu\,,D_\nu]~=~ - g  \mathcal{F}_{\mu\nu}{}^ {M} X_ {M}
~=~ - g  \mathcal{H}_{\mu\nu}{}^ {M} X_ {M}
\;,
\end{equation}
but in the Lagrangian the difference between $\mathcal{F}^M$
and~$\mathcal{H}^M$ is important.

Consistency of the gauge algebra thus requires the introduction of
2-form tensor fields $B_{\mu\nu}{}^ {PN}$. It is important that their
appearance in (\ref{eq:modified-fs}) strongly restricts their possible
representation content. Not only must they transform in the symmetric
product $(NP)$ of the vector field representation as is manifest from
their index structure, but also they appear under contraction with the
tensor $Z^ {M}{}_{ {NP}}$ which in general does not map onto the full
symmetric tensor product in its lower indices, but rather only on a
restricted sub-representation. We will see this explicitly in section
\ref{sec:hierarchyin4d}. It is this sub-representation of ${\rm
  G}$ to which the 2-forms are assigned, and to keep the notation
transparent, we denote the corresponding projector with special
brackets $\llceil{ {NP}}\rrfloor$, such that
\begin{equation}
  \label{eq:brackets}
  Z^ {M}{}_{ {NP}} \,B_{\mu\nu}{}^{NP} ~=~
  Z^ {M}{}_{ {NP}} \,B_{\mu\nu}{}^ {\llceil{ {NP}}\rrfloor}\;,\qquad
  \mbox{etc.}\;.
\end{equation}
The tensor $Z^{M}{}_{{NP}}$ thus plays the role of an {\it
  intertwiner} between vector fields and 2-forms, which encodes the
precise field content of the 2-form tensor fields such that the
consistency of the vector gauge algebra is ensured.

The same pattern continues upon definition of a covariant field
strength for the 2-forms and leads to a hierarchy of $p$-form tensor
fields, which is entirely determined by the choice of the global symmetry
group ${\rm G}$ and its fundamental representation ${\cal R}_{\rm v}$
in which the vector fields transform. In principle this hierarchy
continues indefinitely, but it depends on the actual Lagrangian what
its fate will be. Obviously, $p$ can at most be equal to $d$. When
incorporated into a given Lagrangian the gauge algebra for the
$p$-forms will be deformed and additional structure will appear.  Some
of the $p$-form gauge fields may carry physical degrees of freedom so
they must already be contained in the ungauged Lagrangian, up to
tensor dualities.  For instance, in five dimensions, a vector gauge
field and a 2-form gauge field are dual, so that tensor fields are
potentially present in view of the fact that the ungauged Lagrangian
contains vector fields (which are also essential for the gauging).
This is a generalization of the phenomenon we noted before: in three
dimensions a scalar and a vector field can be dual. Therefore, vector
fields are in principle available as well, as long as the ungauged
Lagrangian contains scalar fields. The construction based on the
Chern-Simons term (c.f. \eqref{eq:CS-E8}) made use of this
observation.

Before discussing these topics any further, let us first turn to a
discussion of the hierarchy for the maximal supergravities. In section
\ref{sec:hierarchyin4d} we will try to further elucidate the structure
of the hierarchy for the case of a generic gauge theory in four
space-time dimensions.

%%%%%%%%%%%%%%%%%%%%%%%%%%%%%%%%%%%%%%%%%%%%%%%%%%%%%%%%%%%%%%%%%
\section{The $p$-form hierarchy for the maximal supergravities}
\setcounter{equation}{0}
\label{sec:p-form-assignments}
%%%%%%%%%%%%%%%%%%%%%%%%%%%%%%%%%%%%%%%%%%%%%%%%%%%%%%%%%%%%%%%%%
The hierarchy of vector and tensor gauge fields that we discussed in
the previous section can be considered for the maximal gauged
supergravities. In that case the gauge group is embedded in the
duality group $\mathrm{G}$, which is known for each space-time
dimension in which the supergravity is defined (see table
\ref{tab:T-tensor-repr}). Once the group $\mathrm{G}$ is specified,
the hierarchy allows a unique determination of the representations of
the higher $p$-forms. Table \ref{tab:vector-tensor-repr} illustrates
this for the maximal supergravities in $d=3,\ldots,7$ space-time
dimensions.  We recall that the analysis described in subsection
\ref{sec:gauge-deformations} did not depend on the number of
space-time dimensions. For instance, it is possible to derive the
representation assignments for $(d\!+\!1)$-rank tensors, although
these do not live in a $d$-dimensional space-time (nevertheless, a
glimpse of their existence occurs in $d$ dimensions via the shift
transformations of the $d$-forms in the general gauged theory, as we
shall see in due course). We also observe that the Hodge duality
between the $p$-form fields that relates the $p$-forms to the
$(d-p-2)$-forms is reflected in table \ref{tab:vector-tensor-repr}, as
the dual form fields appear in conjugate representations of the group
$\mathrm{G}$. This duality depends, of course, sensitively on the
space-time dimension, whereas the only input in the table came from
the duality group and the representations of the low-$p$ form fields.

%%%%%%%%%%%%%%%%%%%%%%%%%%%%%%%%%%%%%%%%%%%%%%%%%%%%%%%%%%%%%%%%%%
%%%%%%%%%%%%%%%%%%%%%%%%%%%%%%%%%%%%%%%%%%%%%%%%%%%%%%%%%%%%%%%%%%
\begin{table}[t]
\centering
\begin{tabular}{l l cccccc  }\hline
~&~&~&~&~&~&~& \\[-4mm]
~ &$~$& 1&2&3&4&5&6  \\   \hline
~&~&~&~\\[-4mm]
7   & ${\rm SL}(5)$ & $\overline{\bf 10}$  & ${\bf 5}$ & $\overline{\bf 5}$ &
${\bf 10}$ &  ${\bf 24}$ & $\overline{\bf 15}+{\bf 40}$  \\[1mm]
6  & ${\rm SO}(5,5)$ & ${\bf 16}_c$ & ${\bf 10}$ & ${\bf 16}_s$ &
  ${\bf 45}$ & ${\bf 144}_s$ &
$\!\!{\bf 10}\!+\! {\bf 126}_s\!+\! {\bf 320}\!\!$\\[.8mm]
5   & ${\rm E}_{6(6)}$ & $\overline{\bf 27}$ & ${\bf 27}$ & ${\bf 78}$
& ${\bf 351}$ & $\!\!{\bf 27}\!+\! {\bf 1728}\!\!$ &  \\[.5mm]
4   & ${\rm E}_{7(7)}$ & ${\bf 56}$ & ${\bf 133}$ & ${\bf 912}$ &
$\!\!\!{\bf 133}\!+\! {\bf 8645}\!\!\!$ &    \\[.5mm]
3   & ${\rm E}_{8(8)}$ & ${\bf 248}$ & ${\bf 1}\!+\! {\bf 3875}$ & ${\bf
  3875}\!+\! {\bf147250}$ & &
\\ \hline
\end{tabular}
%%%%%%%%%%%%%%%%%%%%%%%%%%%%%%%%%%%%%%%%%%%%%%%%%%%%%%%%%%%%%%%%%
\caption{\small
Duality representations of the vector and tensor gauge fields
for gauged  maximal supergravities in space-time dimensions $3\leq
d\leq 7$. The first two columns list the space-time dimension and the
corresponding duality group. }\label{tab:vector-tensor-repr}
\end{table}
%%%%%%%%%%%%%%%%%%%%%%%%%%%%%%%%%%%%%%%%%%%%%%%%%%%%%%%%%%%%%%%%%%
%%%%%%%%%%%%%%%%%%%%%%%%%%%%%%%%%%%%%%%%%%%%%%%%%%%%%%%%%%%%%%%%%%

It is intriguing that the purely group theoretical hierarchy
reproduces the correct assignments consistent with Hodge duality.  In
particular, the assignment of the $(d\!-\!2)$-forms is in line with
tensor-scalar duality, as these forms are dual to the Noether currents
associated with the $\mathrm{G}$ symmetry. In this sense, the duality
group $\mathrm{G}$ implicitly carries information about the space-time
dimension. But the hierarchy naturally extends beyond the
$(d\!-\!2)$-forms and thus to those non-propagating forms whose field
content is not subject to Hodge duality. It is another striking
feature of the hierarchy that the diagonals pertaining to the
$(d\!-\!1)$- and $d$-rank tensor fields refer to the representations
conjugate to those assigned to the embedding tensor and its quadratic
constraint, respectively. This pattern is in fact generic and related
to the special role these forms may play in the
Lagrangian~\cite{deWit:2008ta,deWit:2008gc}. We will briefly discuss
this in the next subsection in a general context. In a later section
\ref{sec:hierarchyin4d} we will illustrate some of this in the context
of a generic gauge theory in four space-time dimensions.

We recall that the embedding tensor is regarded as a so-called {\it
  spurionic} quantity, which transforms under the action of
$\mathrm{G}$, although at the end it will be fixed to a constant
value. For a specific value of the embedding tensor one is describing
a given gauge deformation. Sweeping out the full space of allowed
embedding tensors yields a $\Theta$-independent (and ${\rm
  G}$-covariant) result for the representation of $p$-forms, as is
shown in the table.  This approach shows how the required consistency
under generic gauge deformations imposes strong restrictions on the
field content of the theory.  In the ungauged theory there is a priori
no direct evidence for these restrictions, but in certain cases there
are alternative arguments, based, for instance, on supersymmetry or on
underlying higher-rank Kac-Moody symmetries, which may motivate the
representation content of the $p$-forms. Some of this will be
discussed in subsection \ref{sec:m-theory}.

%%%%%%%%%%%%%%%%%%%%%%%%%%%%%%%%%%%%%%%%%%%%%%%%%%%%%%%%%%
\subsection{More about the hierarchy}
\label{sec:more-about-hierarchy}
%%%%%%%%%%%%%%%%%%%%%%%%%%%%%%%%%%%%%%%%%%%%%%%%%%%%%%%%%%
Now that we have seen some global features of the $p$-form algebra,
let us briefly discuss some more generic aspects. As it turns out the
$p$-forms transform in a sub-representation of the rigid symmetry
group ${\rm G}$ of the theory that is contained in the $p$-fold tensor
product ${\cal R}_{\rm v}^{\otimes p}$, where ${\cal R}_{\rm v}$
denotes the representation of ${\rm G}$ in which the vector fields
transform. In many cases of interest this is the fundamental
representation. We denote these fields by
\begin{eqnarray}
  \label{eq:hierarchy-p}
  \buildrel{\scriptscriptstyle[1]}\over{A}{}^{M\vphantom{]}}\,,\quad
  \buildrel{\scriptscriptstyle[2]}\over{B}{}^{{\llceil MN\rrfloor}}\,,\quad
  \buildrel{\scriptscriptstyle[3]}\over{C}{}^{{\llceil M\llceil
      NP\rrfloor\rrfloor}}\,,\quad
  \buildrel{\scriptscriptstyle[4]}\over{C}{}^{{\llceil M\llceil N\llceil
      PQ\rrfloor\rrfloor\rrfloor}}\,,\quad
  \buildrel{\scriptscriptstyle[5]}\over{C}{}^{{\llceil M\llceil N\llceil
      P\llceil{QR}\rrfloor\cdot\cdot\rrfloor}}\,,\;\; {\rm etc.}\,,
\end{eqnarray}
where we have suppressed space-time indices, and the
special brackets $\llceil\cdots\rrfloor$ are introduced to denote the
relevant sub-representations of ${\cal R}_{\rm v}^{\otimes p}$, just
as was done for $p=2$ in \eqref{eq:brackets}.

Assuming that the theory contains $p$-form fields and that one must
also allow for the presence of the dual $p$-forms, the question arises
what the significance is of the $d$- and $(d\!-\!1)$-forms that appear
in table \ref{tab:vector-tensor-repr}, as those are not dual to any
other forms. In order to explore this further, let us consider the high-$p$
sector of the hierarchy, starting with the $(d-3)$- and the
$(d-2)$-forms, transforming in the representations that are conjugate
to those assigned to the 1- and the 2-forms \cite{deWit:2008gc}. In
view of the fact that the theory is invariant under the group
$\mathrm{G}$ prior to switching on the gauge couplings, there exists a
set of conserved 1-forms given by the Noether currents, transforming
in the adjoint representation, which are dual to the $(d\!-\!2)$-forms.
Hence we expect these forms to belong to the adjoint representation.
Furthermore we expect $(d\!-\!3)$-forms that are dual to the vector
fields, to transform in the ${\rm G}$-representation ${\cal R}_{\rm
  v*}$, which is conjugate to the vector field representation ${\cal
  R}_{\rm v}$. When considering these high-rank $p$-forms it is
convenient to switch to a notation adapted to this particular field
content and to identify the $(d\!-\!3)$- and $(d\!-\!2)$-forms as,
\begin{eqnarray}
  \label{eq:d-3-d-2}
  \buildrel{\scriptscriptstyle[d-3]}\over{C}{}^{ {M}_1\llceil
   {M}_2\llceil\cdots {M}_{d-3}\rrfloor\cdot\cdot\rrfloor}
   &\sim & \buildrel{\scriptscriptstyle[d-3]}\over{C}_ M
   \;, \nonumber\\[.2ex]
   \buildrel{\scriptscriptstyle[d-2]}\over{C}{}^{ {M}_1\llceil
   {M}_2\llceil\cdots {M}_{d-2}\rrfloor\cdot\cdot\rrfloor}
   &\sim & \buildrel{\scriptscriptstyle[d-2]}\over{C}_\alpha \;.
\end{eqnarray}
We may then explicitly study the end of the $p$-form hierarchy by
imposing the general structure in a schematic form,
\begin{eqnarray}
  \label{eq:general-hierarchy-end}
  \delta \buildrel{\scriptscriptstyle[d-3]}\over{C}{}_M &=&
  (d-3) \,{\mathrm D}\!
  \buildrel{\scriptscriptstyle[d-4]}\over{\Phi}{}_M + \cdots
  - g\,Y_M{}^\alpha\,\buildrel{\scriptscriptstyle[d-3]}\over{\Phi}{}_\alpha
  \;,\nonumber\\[1.5ex]
  \delta \buildrel{\scriptscriptstyle[d-2]}\over{C}{}_\alpha &=&
  (d-2)\,{\mathrm D}\!
  \buildrel{\scriptscriptstyle[d-3]}\over{\Phi}{}_\alpha + \cdots
  - g \,Y_{\alpha,M}{}^\beta
  \buildrel{\scriptscriptstyle[d-2]}\over{\Phi}{}^M{}_\beta
  \;,\nonumber\\[1.5ex]
  \delta \buildrel{\scriptscriptstyle[d-1]}\over{C}{}^M{}_\alpha &=&
  (d-1)\,{\mathrm D}\!
  \buildrel{\scriptscriptstyle[d-2]}\over{\Phi}{}^M{}_\alpha+ \cdots
  - g\,Y^M{}_{\alpha, PQ}{}^\beta
  \buildrel{\scriptscriptstyle[d-1]}\over{\Phi}{}^{PQ}{}_\beta
  \;,\nonumber\\[1.5ex]
  \delta \buildrel{\scriptscriptstyle[d]}\over{C}{}^{MN}{}_\alpha &=&
  d\,{\mathrm D}\!
  \buildrel{\scriptscriptstyle[d-1]}\over{\Phi}{}^{MN}{}_\alpha
  + \cdots - g\,Y^{MN}{}_{\alpha, PQR}{}^\beta \,
  \buildrel{\scriptscriptstyle[d]}\over{\Phi}{}^{PQR}{}_\beta
  \;,\nonumber\\[1.5ex]
  \delta \buildrel{\scriptscriptstyle[d+1]}\over{C}{}^{PQR}{}_\alpha &=&
  (d+1)\,{\mathrm D}\!
  \buildrel{\scriptscriptstyle[d]}\over{\Phi}{}^{PQR}{}_\alpha
  + \cdots  \,,
\end{eqnarray}
where we have indicated the most characteristic terms in the $p$-form
transformations. We included the transformations associated to the
$(d\!+\!1)$-form. Although this form does not exist in $d$ dimensions,
its associated gauge transformations still play a role as they act on
the $d$-form gauge fields, as is exhibited above.

The above schematic result \eqref{eq:general-hierarchy-end}, as well
as the rest of the hierarchy, contains several new intertwining
tensors $Y$ that connect the representations associated with two
successive form fields. Formally they may be considered as a map
\begin{eqnarray}
Y^{[p]}:\,
{\cal R}_{\rm v}^{\otimes (p+1)}\;\longrightarrow\;{\cal R}_{\rm v}^{\otimes p}
\;,
  \label{eq:map}
\end{eqnarray}
which has a non-trivial kernel whose complement defines the
representation content of the $(p+1)$-forms that is required for
consistency of the deformed $p$-form gauge algebra. The lowest-rank
intertwining tensors are given by
\begin{eqnarray}
  \label{eq:lowerY}
Y^{[0]}:\,
{\cal R}_{\rm v}\;\longrightarrow\;{\cal R}_{\rm adj}
\;,\qquad
Y^{[1]}:\,
{\cal R}_{\rm v}^{\otimes 2}\;\longrightarrow\;{\cal R}_{\rm v}
\;,
\end{eqnarray}
corresponding to $p=0,1$, with
$(Y^{[0]})^\alpha{}_{M}=\Theta_M{}^\alpha$ and
$(Y^{[1]})^M{}_{PQ}=Z^M{}_{PQ}$. For higher $p$, the intertwining
tensors can be defined recursively, as was demonstrated in
\cite{deWit:2008ta}. All intertwining tensors are proportional to
the embedding tensor and they must be mutually orthogonal,
\begin{eqnarray}
  \label{eq:orthoY1}
  Y^{[p]}  \cdot Y^{[p+1]} &\approx& 0
  \;.
\end{eqnarray}
This is a generalization of \eqref{eq:closure}.  More explicitly, the
orthogonality relations read,
\begin{eqnarray}
  \label{eq:orthoY}
  Y^{{K}_2\llceil{K}_3\llceil\cdots {K}_p\rrfloor
  \cdot\cdot\rrfloor}
  {}_{{M}_1\llceil {M}_2\llceil\cdots
   {M}_p\rrfloor\cdot\cdot\rrfloor} \;
   Y^{{M}_1\llceil {M}_2\llceil\cdots
   {M}_p\rrfloor\cdot\cdot\rrfloor}
  {}_{{N}_0\llceil{N}_1\llceil\cdots
   {N}_p\rrfloor\cdot\cdot\rrfloor}
   &\approx&{} 0 \;,
\end{eqnarray}
where `weakly zero' indicates that the expression vanishes as a
consequence of the quadratic constraint (\ref{eq:gauge-inv-embedding})
on the embedding tensor.

It can be shown that the intertwining tensors appearing in
\eqref{eq:general-hierarchy-end} take the form,
\cite{deWit:2008gc},
\begin{eqnarray}
  \label{eq:interpolaters}
  Y_M{}^\alpha &=&
  \Theta_M{}^\alpha \;,
  \nonumber\\[1.5ex]
  Y_{\alpha,M}{}^\beta &=&
  \delta_\alpha \Theta_M{}^\beta\;,
  \nonumber\\[1.5ex]
  Y^M{}_{\alpha, PQ}{}^\beta  &=&{}-
  \frac{\partial{\cal Q}_{PQ}{}^\beta}{\partial\,\Theta_M{}^\alpha} \;,
  \nonumber\\[1.5ex]
  Y^{MN}{}_{\alpha, PQR}{}^\beta  &=&
  -\delta^M_P\,Y^N{}_{\alpha, QR}{}^\beta
  - X_{PQ}{}^M\,\delta_R^N\delta_\alpha^\beta
  - X_{PR}{}^N\,\delta_Q^M \delta_\alpha^\beta
  + X_{P\alpha}{}^\beta\,\delta_R^N\delta_Q^M
  \;.
\end{eqnarray}
It is straightforward to verify that these intertwining tensors satisfy
the mutual orthogonality property (\ref{eq:orthoY}), and one easily
derives,
\begin{eqnarray}
  \label{eq:YY-aprox-Q}
  Y_M{}^\alpha \,Y_{\alpha,N}{}^\beta &\approx&0  \;,
  \nonumber\\[.8ex]
  Y_{\alpha,N}{}^\beta\,Y^N{}_{\beta, PQ}{}^\gamma &\approx& 0
  \;,\nonumber \\[.8ex]
  Y^M{}_{\alpha, KL}{}^\beta\,
  Y^{KL}{}_{\beta, PQR}{}^\gamma   &\approx&0  \;.
\end{eqnarray}
There is an additional identity which holds identically, without
making reference to the constraint \eqref{eq:constraint2},
\begin{equation}
  \label{eq:Y-Q-vanish}
  Y^{MN}{}_{\alpha, PQR}{}^\beta \; {\cal Q}_{MN}{}^\alpha
  = 0 \;.
\end{equation}
The relevance of this result will be discussed below.

From~(\ref{eq:interpolaters}) we can now directly read off the
representation content of the $(d\!-\!1)$- and the $d$-forms that
follows from the hierarchy: the form of $Y_{\alpha,M}{}^\beta$ and
$Y^M{}_{\alpha, PQ}{}^\beta$ shows that these forms transform in the
representations dual to the embedding tensor $\Theta_M{}^\beta$ and
the quadratic constraint ${\cal Q}_{PQ}{}^\beta$, respectively.  As
such, they can naturally be coupled, acting as Lagrange multipliers
enforcing the property that the embedding tensor is space-time
independent and gauge invariant \cite{deWit:2008ta}.  This idea has
been worked out explicitly in the context of maximal supergravity in
three space-time dimensions, and subsequently it has been argued that
this situation can also be realized in a more general context
\cite{deWit:2008gc}. Hence we view the embedding tensor as a
space-time dependent scalar field, transforming in the
$\mathrm{G}$-representation that is allowed by the representation
constraint. To the original Lagrangian $\mathcal{L}_0$ which may
depend on $p$-forms with $p\leq d-2$, we then add the following
terms,
\begin{equation}
  \label{eq:addtion}
  \mathcal{L} = \mathcal{L}_0+\mathcal{L}_\mathrm{C}\,,
\end{equation}
with
\begin{eqnarray}
  \label{eq:LC}
  \mathcal{L}_\mathrm{C} &\propto&{}
  \varepsilon^{\mu_1\cdots\mu_d}\left\{ d\,g \,
  C_{\mu_2\cdots\mu_d}{}^{M}{}_\alpha \,D_{\mu_1}\Theta_M{}^\alpha
  + g^{2}\,
  C_{\mu_1\cdots\mu_d}{}^{MN}{}_\alpha\;
  \mathcal{Q}_{MN}{}^\alpha\right\}  \;.
\end{eqnarray}
Note that the identity (\ref{eq:Y-Q-vanish}) ensures that this
Lagrangian is invariant under the shift transformation of the $d$-rank
tensor field. Variation of this Lagrangian with
respect to $\Theta_M{}^\alpha(x)$ leads to the following expression,
\begin{eqnarray}
  \label{eq:delta-LC}
  \delta\mathcal{L}_\mathrm{C} &\propto& -
  g\, \varepsilon^{\mu_1\cdots\mu_d} \;\delta\Theta_M{}^\alpha
  \nonumber \\[.5ex]
  &&{} \quad\times
  \Big[d\,D_{\mu_1}C_{\mu_2\cdots\mu_d}{}^M{}_\alpha  +g\,
  Y^M{}_{\alpha,PQ}{}^\beta\,
  C_{\mu_1\cdots\mu_d}{}^{PQ}{}_\beta
  +d\, g\, A_{\mu_1}\, Y_{\alpha,N}{}^\beta\;
  C_{\mu_2\cdots\mu_d}{}^N{}_\beta \Big]
  \,.\nonumber\\[.5ex]
 \end{eqnarray}
This result can be written as follows,
\begin{equation}
  \label{eq:vra-into-H}
  \delta\mathcal{L}_\mathrm{C}\propto -
  g\,\varepsilon^{\mu_1\cdots\mu_d}\,\Big[
  \mathcal{H}_{\mu_1\cdots\mu_d}{}^M{}_\alpha + d\,  A_{[\mu_1}{}^M
  \;\mathcal{H}_{\mu_2\cdots\mu_d]\;\alpha} +\cdots\Big]
  \,\delta\Theta_M{}^\alpha \,,
\end{equation}
by including unspecified terms involving form fields of rank $p\leq
d-2$. These terms are expected to arise from the $\Theta$-variation of
the Lagrangian $\mathcal{L}_0$, but they cannot be evaluated in full
generality as this depends on the details of the latter Lagrangian. We
return to this issue in the next section where we study the situation
in four space-time dimensions.

%%%%%%%%%%%%%%%%%%%%%%%%%%%%%%%%%%%%%%%%%%%%%%%%%%%%
\subsection{M-Theory}
\label{sec:m-theory}
%%%%%%%%%%%%%%%%%%%%%%%%%%%%%%%%%%%%%%%%%%%%%%%%%%%%
It is an obvious question whether the systematic features shown in
table \ref{tab:vector-tensor-repr} have a natural explanation in terms
of M-theory. Supergravity may already contain some of the fields that
carry charges that are required for some of these gaugings.  Indeed,
we already noted in subsection \ref{sec:hidden-symmetries} that the
towers of massive Kaluza-Klein states carry charges that couple to the
Kaluza-Klein gauge fields emerging from the higher-dimensional metric.
This is of direct relevance to the so-called Scherk-Schwarz reductions
\cite{Scherk:1979zr}.  However, these Kaluza-Klein states cannot
generally be assigned to representations of the duality group and
therefore there must be extra degrees of freedom whose origin cannot
be understood within the context of a dimensional compactification of
supergravity.\footnote{%%%%%%%%%%%%%%%%%%%%%%%%%%%%%%%%%%%%%%%%%%
  In view of the fact that the Kaluza-Klein states are 1/2-BPS, also
  these extra degrees of freedom must correspond to 1/2-BPS states. %
} %%%%%%%%%%%%%%%%%%%%%%%%%%%%%%%%%%%%%%%%%%%%%%%%%%
This phenomenon was discussed some time ago, for
instance, in \cite{Obers:1998fb,deWit:2000wu}.

General gaugings of maximal supergravity constructed in recent years
obviously extend beyond gaugings whose charges can be fully understood
from supergravity degrees of freedom in higher dimensions. The duality
covariant embedding tensor encodes all the possible charges which must
somehow have their origin in M-theory. Indeed there are indications
that this is the case. In this way the gauging acts as a probe of
M-theory degrees of freedom.

Independent evidence that this relation with M-theory degrees of
freedom is realized, is provided by the work of \cite{Elitzur:1997zn}
(see also, \cite{Obers:1998fb} and references quoted therein) where
matrix theory \cite{deWit:1988ig,Banks:1996vh} is considered in a
toroidal compactification.  This work is based on the correspondence
between $N=4$ super-Yang-Mills theory in $n+1$ dimensions ($n\leq9$),
on a (rectangular) spatial torus $\tilde{T}^n$ with radii $s_1,\ldots,
s_n$, and M-theory in the infinite-momentum frame on the dual torus
$T^n$ with radii $R_1,R_2, \ldots, R_n$, where $s_i=
l_\mathrm{p}^3/R_{11}R_i$. Here $l_\mathrm{p}$ denotes the Planck
length in eleven dimensions and $R_{11}$ is the length of the
compactified eleventh dimension. The latter dimension, together with
the time dimension and the spatial dimensions that do not belong to
$T^n$ constitute the $d$-dimensional space-time that is relevant in
the comparison. Just as before $d=11-n$.

The conjecture is that M-theory should be invariant under both the
permutations of the radii $R_i$ and under T-duality of type-IIA string
theory. The relevant T-duality transformations follow from making two
consecutive T-dualities on two different circles. When combined with
the permutation symmetry, T-duality can be represented by
($i\not=j\not=k\not=i$)
\begin{eqnarray}
  \label{eq:T-duality}
  R_i\to \frac{l_\mathrm{p}^3}{R_j R_k} \;, \quad
  R_j\to \frac{l_\mathrm{p}^3}{R_k R_i} \;, \quad
  R_k\to \frac{l_\mathrm{p}^3}{R_i R_j} \;, \quad
  l_\mathrm{p}^3 \to \frac{l_\mathrm{p}^6}{R_i R_j R_k} \;.
\end{eqnarray}

The above transformations generate a discrete group which coincides
with the Weyl group of $\mathrm{E}_n$; on the Yang-Mills side, the
elementary Weyl reflections correspond to permutations of the
compactified coordinates (generating the Weyl group of
$\mathrm{SL}(n)$) and to Montonen-Olive duality
$g_\mathrm{eff}\rightarrow 1/g_\mathrm{eff}$ (corresponding to
reflections with respect to the exceptional node of the $\mathrm{E}_n$
Dynkin diagram).  This Weyl group, which leaves the rectangular shape
of the compactification torus invariant, can be realized as a discrete
subgroup of the compact subgroup of $\mathrm{E}_{n(n)}$, and
consequently as a subgroup of the conjectured non-perturbative duality
group $\mathrm{E}_{n(n)}(\mathbb{Z})$ \cite{Hull:1994ys}.
Representations of this symmetry can now be generated by mapping out
the Weyl orbits starting from certain states. For instance, one may
start with Kaluza-Klein states on $T^n$, whose masses are proportional
to $M\sim 1/R_i$. The action of the Weyl group then generates new
states, such as the ones that can be identified with two-branes
wrapped around the torus, whose masses are of order $M\sim
R_jR_k/l_\mathrm{p}^3$, and so on. To be specific, let us consider the
situation for $n=4$ and $d=7$ and start from the four Kaluza-Klein
states with masses $M\sim 1/R_i$, where $i=1,2,3,4$. Upon the action
of \eqref{eq:T-duality}, we find six two-brane states wrapped on
$T^4$.  Repeated application of \eqref{eq:T-duality} does not give
rise to new states, so that we find precisely ten particle states
(i.e., massive charged particle states from the seven-dimensional
perspective):
\begin{equation}
  \label{eq:particle-T4}
  \mbox{10 particle states} \; \left\{
  \begin{array}{lcl}
     \mbox{4 KK states on $T^4$}  &:& M\sim \displaystyle{
     \frac{1}{R_i}} \\[5mm]
     \mbox{6 two-brane states wrapped on $T^4$}  &:& M\sim \displaystyle{
     \frac{R_jR_k}{l_\mathrm{p}^3}}
    \end{array}  \right.
\end{equation}
Here $j\not=k$. The pointlike charges associated with these states can
couple to ten gauge fields, and this is precisely the number of
1-forms in table \ref{tab:vector-tensor-repr} for $d=7$.

Likewise we can consider 2-brane states wrapped on $T^4\times
\mathbb{R}^{11}$, where $\mathbb{R}^{11}$ denotes the eleventh
dimension, which has been compactified to size $R_{11}$. There
are four such states with masses $M\sim R_{11}R_i/l_\mathrm{p}^3$.
Application of \eqref{eq:T-duality} now leads to only one additional
state, corresponding to a five-brane wrapped on $T^4\times
\mathbb{R}^{11}$. Hence we find five stringlike states, from the
perspective of the seven-dimensional space-time,
\begin{equation}
  \label{eq:string-T4}
  \mbox{5 string states} \; \left\{
  \begin{array}{lcl}
     \mbox{4 two-brane states wrapped on $T^4\times\mathbb{R}^{11}$}
     &:& M\sim \displaystyle{
     \frac{R_{11} R_i}{l_\mathrm{p}^3}} \\[5mm]
     \mbox{1 five-brane state wrapped on $T^4\times\mathbb{R}^{11}$}
     &:& M\sim \displaystyle{
     \frac{R_{11}R_1R_2R_3R_4}{l_\mathrm{p}^6}}
    \end{array}  \right.
\end{equation}
Altogether we thus have a multiplet consisting of five different
string states, which can couple to five different 2-form fields. This
is precisely the number of 2-forms listed in table
\ref{tab:vector-tensor-repr} for $d=7$.

Similar arguments apply to the other states, except that when the
representation has weights of different lengths, one needs several
different Weyl orbits to recover all states in the representation.
For instance, there are only 2160 states for $\mathrm{E}_{8(8)}$,
which must be supplemented by eight-brane states to obtain the full
$\bf{3875}$ representation of $\mathrm{E}_{8(8)}$.  In this way one
obtains complete multiplets of the duality group (taking into account
that some states belonging to the representation will vanish under the
Weyl group and will therefore remain inaccessible by this
construction).

The representations in the table were also found in
\cite{Iqbal:2001ye}, where a `mysterious duality' was exhibited
between toroidal compactifications of M-theory and del Pezzo surfaces.
Here the M-theory dualities are related to global diffeomorphisms that
preserve the canonical class of the del Pezzo surface. Again the
representations thus found are in good agreement with the
representations in table~\ref{tab:vector-tensor-repr}.

For $n=9$, the multiplets given in \cite{Elitzur:1997zn} have
infinitely many components. Indeed, there are hints that the above
considerations concerning new M-theoretic degrees of freedom can be
extended to infinite-dimensional duality groups. Already some time ago
\cite{West:2004kb} it was shown from an analysis of the indefinite
Kac--Moody algebra $\mathrm{E}_{11}$ that the decomposition of its
so-called L1 representation at low levels under its finite-dimensional
subalgebra $\mathrm{SL}(3) \times \mathrm{E}_{8}$ yields the same
$\mathbf{3875}$ representation that appears for the two-forms shown in
table~\ref{tab:vector-tensor-repr}. This analysis has meanwhile been
extended
\cite{Riccioni:2007au,Bergshoeff:2007qi,Riccioni:2007ni,Bergshoeff:2008qd}
to other space-time dimensions and higher-rank forms, and again there
is a clear overlap with the representations in
table~\ref{tab:vector-tensor-repr}.  Nevertheless it remains far from
clear what all these (infinitely many) new degrees of freedom would
correspond to, and how they would be concretely realized. Concerning
the physical interpretation of the new states, a first step was taken
in \cite{Englert:2007qb}, where an infinite multiplet of BPS states is
generated from the M2 brane and M5 brane solutions of $D=11$
supergravity by the iterated action of certain $A_1^{(1)}$ subgroups
of the $\mathrm{E}_9$ Weyl group. For more recent work, see
\cite{Bergshoeff:2008xv}. In the context of gauged supergravities, the
significance of these states may become clearer with the exploration
of maximal gauged supergravities in {\it two} space-time dimensions
\cite{Samtleben:2007an}, where the embedding tensor transforms in the
so-called basic representation of $\mathrm{E}_9$ (which is infinite
dimensional).

%%%%%%%%%%%%%%%%%%%%%%%%%%%%%%%%%%%%%%%%%%%%%%%%%%%%%%%%%%%
%%%%%%%%%%%%%%%%%%%%%%%%%%%%%%%%%%%%%%%%%%%%%%%%%%%%%%%%%%%
\section{The $p$-form hierarchy in four space-time dimensions}
\setcounter{equation}{0}
\label{sec:hierarchyin4d}
%%%%%%%%%%%%%%%%%%%%%%%%%%%%%%%%%%%%%%%%%%%%%%%%%%%%%%%%%%%%
In this section we present the $p$-form hierarchy for a generic $d=4$
dimensional gauge theory, following earlier work in
\cite{deWit:2005ub,Schon:2006kz,Derendinger:2007xp,deVroome:2007zd}.
Although matters will become more complicated towards the end, we hope
that this illustrates a number of characteristic features. First of
all, we will see that the $p$-form fields belong to restricted
representations, as was noted previously. Then we will exhibit the
fact that the gauge transformations of the hierarchy are deformed when
considered in the context of a specific Lagrangian, rather than as an
abstract algebra. And finally, we will be more explicit (although we
will refrain from giving all the details) about the introduction of
the 3- and 4-form fields.  For simplicity we suppress the
gravitational interactions and consider a Lagrangian depending on $n$
abelian gauge fields $A_\mu{}^\Lambda$ (so that $\Lambda= 1,\ldots,n$,
where $n$ has no relation to the torus dimension, as in the
previous sections). We start without charged fields so that the gauge
fields $A_\mu{}^\Lambda$ appear exclusively through the field
strengths, ${F}_{\mu\nu}{}^\Lambda =
2\,\partial_{[\mu}A_{\nu]}{}^\Lambda$.

The field equations for these fields and the Bianchi identities for
the field strengths comprise $2n$ equations,
\begin{equation}
  \label{eq:eom-bianchi}
  \partial_{[\mu} {F}_{\nu\rho]}{}^\Lambda
   = 0 = \partial_{[\mu} {G}_{\nu\rho]\,\Lambda} \,,
\end{equation}
where
\begin{equation}
  \label{eq:def-G}
  {G}_{\mu\nu\,\Lambda} =
  \varepsilon_{\mu\nu\rho\sigma}\,
  \frac{\partial \mathcal{L}}{\partial{F}_{\rho\sigma}{}^\Lambda}
  \;,
\end{equation}
where we use a metric with signature $(-,+,+,+)$ and
$\varepsilon_{0123}=1$ denotes the four-dimensional
Levi-Civita symbol in four Minkowskian dimensions.

It is convenient to combine the tensors $F_{\mu\nu}{}^\Lambda$ and
$G_{\mu\nu\Lambda}$ into a $2n$-dimensional vector,
\begin{equation}
  \label{eq:GM}
  G_{\mu\nu}{}^M= \begin{pmatrix} {F}_{\mu\nu}{}^\Lambda\\[1.5mm]
  {G}_{\mu\nu\Lambda}
\end{pmatrix} \,,
\end{equation}
so that (\ref{eq:eom-bianchi}) reads $\partial_{[\mu}
{G}_{\nu\rho]}{}^M = 0$. Obviously, these $2n$ equations are invariant
under real $2n$-dimensional rotations of the tensors $G_{\mu\nu}{}^M$,
\begin{equation}
  \label{eq:em-duality}
  \begin{pmatrix}{F}^\Lambda\\[1.5mm] {G}_\Lambda\end{pmatrix}
  \longrightarrow
  \begin{pmatrix}{\tilde F}^\Lambda\\[1.5mm]
  {\tilde G}_\Lambda  \end{pmatrix} =
  \begin{pmatrix}U^\Lambda{}_\Sigma & Z^{\Lambda\Sigma} \\[1.5mm]
    W_{\Lambda\Sigma} & V_\Lambda{}^\Sigma
  \end{pmatrix}
  \begin{pmatrix}{F}^\Sigma\\[1.5mm]  {G}_\Sigma
  \end{pmatrix} \,.
\end{equation}
The first half of the rotated tensors can be adopted as new field
strengths defined in terms of new gauge fields, and constraints on the
remaining tensors can then be interpreted as field equations belonging
to some new Lagrangian $\tilde{\mathcal{L}}$ expressed in terms of the
new field strengths $\tilde F_{\mu\nu}{}^\Lambda$, with $\tilde
G_{\mu\nu}=\varepsilon_{\mu\nu\rho\sigma} \partial
\tilde{\mathcal{L}}/\partial \tilde F_{\rho\sigma}{}^\Lambda$. In
order that such a Lagrangian exists, the real matrix in
(\ref{eq:em-duality}) must belong to the group ${\rm
  Sp}(2n;\mathbb{R})$. This group consists of real matrices that leave
the skew-symmetric tensor $\Omega_{MN}$ invariant,
\begin{equation}
  \label{eq:omega}
  \Omega = \left( \begin{array}{cc}
0 & {\bf 1}\\ \!-{\bf 1} & 0
\end{array} \right)  \;.
\end{equation}
The conjugate matrix $\Omega^{MN}$ is defined by
$\Omega^{MN}\Omega_{NP}= - \delta^M{}_{\!\!P}$.\footnote{%%%%%%%%%%%%%%%%%%
  Here we employ an $\mathrm{Sp}(2n,\mathbb{R})$ covariant notation
  for the $2n$-dimensional symplectic indices $M,N,\ldots$, such that
  $Z^M= (Z^\Lambda, Z_\Lambda)$. Likewise we use vectors with lower
  indices according to $Y_M= (Y_\Lambda,Y^\Lambda)$, transforming
  according to the conjugate representation so that $Z^M\,Y_M$ is
  invariant under $\mathrm{Sp}(2n;\mathbb{R})$. }
%%%%%%%%%%%%%%%%%%%%%%%%%%%%%%%%%%%%%%%%%%%%%%%%%%%%%%%%%%%%%%%%%%%%%
The ${\rm Sp}(2n;\mathbb{R})$ transformations are known as
electric/magnetic dualities, which also act on electric and magnetic
charges (for a review of electric/magnetic duality, see
\cite{deWit:2001pz}). The Lagrangian depends on the electric/magnetic
duality frame and is therefore not unique. Different Lagrangians
related by electric/magnetic duality lead to equivalent field
equations and Bianchi identities, and thus belong to the same
equivalence class. Since the relationship \eqref{eq:em-duality}
between the old and the new field strengths is not a local one, the
new Lagrangian can in general not be obtained by straightforward
substitution.  Instead one may derive,
\begin{equation}
  \label{eq:newlagrangian}
  \tilde{\mathcal{L}}(\tilde F) + \ft18
  \varepsilon^{\mu\nu\rho\sigma} \tilde F_{\mu\nu}{}^\Lambda
  \, \tilde G_{\rho\sigma\Lambda}
  = \mathcal{L}(F) +\ft18 \varepsilon^{\mu\nu\rho\sigma}
  F_{\mu\nu}{}^\Lambda \, G_{\rho\sigma \Lambda} \,,
\end{equation}
up to terms independent of $F^{\mu\nu}{}^\Lambda$. Clearly the
Lagrangian does not transform as a function, since
\begin{equation}
  \label{eq:not-function}
\tilde{\mathcal{L}}(\tilde F) \not= {\mathcal{L}} (F)\,,
\end{equation}
but the combination,
\begin{equation}
  \label{eq:FG-function}
\mathcal{L}(F) +\ft18 \varepsilon^{\mu\nu\rho\sigma}
  F_{\mu\nu}{}^\Lambda \, G_{\rho\sigma \Lambda} \,,
\end{equation}
does, in view of \eqref{eq:newlagrangian}.\footnote{%%%%%%%%%%%%%%%
  When the field equations of the vector fields are imposed, the
  Lagrangian does in fact transform as a function under
  electric/magnetic duality. }
%%%%%%%%%%%%%%%%%%%%%%%%%%%%%%%%%%%%%%%%%%%%%%%%%%%%%%%%%%

When $\mathcal{L}$ remains unchanged under the duality transformation,
{\it i.e.} when
\begin{equation}
  \label{eq:dual-invariant}
  \tilde{\mathcal{L}}(\tilde F) = \mathcal{L}(\tilde F)\,,
\end{equation}
then the theory is {\it invariant} under the corresponding
transformations. It is usually difficult to verify this equation
explicitly. Instead one may verify that the substitution
$F_{\mu\nu}{}^\Lambda\to \tilde F_{\mu\nu}{}^\Lambda$ into the
derivatives $\partial{\mathcal{L}}(F)/\partial F^{\mu\nu}{}^\Lambda$
correctly induces the symplectic transformations of the field
strengths $G_{\mu\nu \Lambda}$. In this case, the linear combination
\eqref{eq:FG-function}, which can also be written as $\mathcal{L}(F)-
\tfrac12 F_{\mu\nu}{}^\Lambda \,\partial\mathcal{L}(F)/\partial
F_{\mu\nu}{}^\Lambda$, must be an {\it invariant} function under
$F_{\mu\nu}{}^\Lambda \to \tilde F_{\mu\nu}{}^\Lambda $.  Note that in
the literature the word duality is used both for equivalence and for
invariance transformations. But the duality group $\mathrm{G}$
introduced before, includes only those $\mathrm{Sp}(2n;\mathbb{R})$
transformations that satisfy \eqref{eq:dual-invariant}.

For clarity we first consider a sub-class of the duality
transformations consisting of those transformations
\eqref{eq:em-duality} for which $Z=0$.  These are the transformations
that act locally on the various fields, as can be seen from the fact
that $\tilde F_{\mu\nu}^\Lambda=
U^\Lambda{}_\Sigma\,F_{\mu\nu}{}^\Sigma$. Because the transformation
must belong to $\mathrm{Sp}(2n;\mathbb{R})$, it follows
$UV^\mathrm{T}= \mathbf{1}$ and that $U^\mathrm{T} W$ is a symmetric
matrix. Using \eqref{eq:newlagrangian} and \eqref{eq:dual-invariant},
one easily derives that the Lagrangian changes by a total derivative,
\begin{equation}
  \label{eq:variationPQ}
  \mathcal{L}(U^\Lambda{}_\Sigma \,F^\Sigma) = \mathcal{L}(F^\Lambda)
  -\ft18 \varepsilon^{\mu\nu\rho\sigma}
  (U^\mathrm{T} W)_{\Lambda\Sigma}\,
  F_{\mu\nu}{}^\Lambda F_{\rho\sigma}{}^\Sigma\,.
\end{equation}
So far we have only indicated the dependence on the field strengths
$F_{\mu\nu}{}^\Lambda$, but other fields may be present as well and
will transform locally among themselves. Their transformations have to
be included in \eqref{eq:dual-invariant}.

We now consider the introduction of a non-abelian gauge group that
will act non-trivially on the vector fields and must therefore involve
a subgroup of the duality group. Because the duality group acts both on
electric and on magnetic charges, in view of the fact that it mixes
field strengths with dual field strengths as shown by
(\ref{eq:em-duality}), we must eventually introduce magnetic gauge
fields $A_{\mu \Lambda}$ as well, following the procedure explained in
\cite{deWit:2005ub}. The $2n$ gauge fields $A_\mu{}^M$ will then
comprise both type of fields, $A_\mu{}^M= (A_\mu{}^\Lambda,
A_{\mu\Lambda})$. The role played by the magnetic gauge fields will be
clarified later. For the moment one may associate $A_{\mu\,\Lambda}$
with the dual field strengths ${G}_{\mu\nu \,\Lambda}$, by writing
${G}_{\mu\nu\,\Lambda} \equiv 2 \,\partial_{[\mu} A_{\nu]\Lambda}$.

The gauge group generators (as far as their embedding in the duality
group is concerned) are then defined as follows. The generators of the
subgroup that is gauged, are $2n$-by-$2n$ matrices $X_M$, where we are
assuming the presence of both electric and magnetic gauge fields, so
that the generators decompose according to
$X_M=(X_{\Lambda},X^\Lambda)$.  Obviously $X_{\Lambda N}{}^P$ and
$X^\Lambda{}_N{}^P$ can be decomposed into the generators of the
duality group and are thus consistent with the infinitesimal form of
the transformations (\ref{eq:em-duality}). Denoting the gauge group
parameters by $\Lambda^M(x) = (\Lambda^\Lambda(x),
\Lambda_\Lambda(x))$, $2n$-dimensional ${\rm Sp}(2n;\mathbb{R})$
vectors $Y^M$ and $Z_M$ transform according to
\begin{equation}
  \label{eq:gauge-tr-Y-Z}
  \delta Y^M = -g \Lambda^N \,X_{NP}{}^M \,Y^P\,,\qquad
    \delta Z_M = g \Lambda^N \,X_{NM}{}^P \,Z_P\,,
\end{equation}
where $g$ denotes a universal gauge coupling constant. Covariant
derivatives thus take the form,
\begin{eqnarray}
  \label{eq:cov-derivative}
  D_\mu Y^M &=& \partial_\mu Y^M + g A_\mu{}^N\, X_{NP}{}^M\,Y^P
  \nonumber \\
  &=& \partial_\mu Y^M + g A_\mu{}^\Lambda\, X_{\Lambda P}{}^M\,Y^P +
  g A_{\mu\Lambda}\, X^\Lambda{}_{P}{}^M\,Y^P \,,
\end{eqnarray}
and similarly for $D_\mu Z_M$. The gauge fields then transform
according to
\begin{equation}
  \label{eq:gauge-tr-A}
  \delta A_\mu{}^M = D_\mu \Lambda^M= \partial_\mu \Lambda^M + g\,
  X_{PQ}{}^M  A_\mu{}^P\, \Lambda^Q \,.
\end{equation}

After replacing ordinary by covariant derivatives and field strengths,
the Lagrangian is in general not invariant. To see this, let us
consider gauge transformations belonging to the subgroup considered
earlier, for which the field transformations take a local form. Hence
we set $X^\Lambda{}_N{}^P=0$ so that the magnetic gauge fields will
not enter and $X_\Lambda{}^{\Sigma\Gamma}= 0$. In this case we can
make use of the result \eqref{eq:variationPQ}. As before the
Lagrangian is not invariant and it changes with the covariantized
form of the variation (\ref{eq:variationPQ}),
\begin{equation}
  \label{eq:gauge-variationPQ}
  \mathcal{L} \to  \mathcal{L} + \ft18 \,
  \varepsilon^{\mu\nu\rho\sigma}  \,\Lambda^\Lambda\,
  X_{\Lambda\Sigma\Gamma}\,
  \mathcal{F}_{\mu\nu}{}^\Sigma \mathcal{F}_{\rho\sigma}{}^\Gamma\,,
\end{equation}
where the tensors $\mathcal{F}_{\mu\nu}{}^\Lambda$ denote the
non-abelian field strengths,
\begin{equation}
  \label{eq:nonabelian-FS}
  \mathcal{F}_{\mu\nu}{}^\Lambda = \partial_\mu A_\nu{}^\Lambda -
  \partial_\nu A_\mu{}^\Lambda  + g\, X_{\Sigma\Gamma}{}^\Lambda\,
  A_{[\mu}{}^\Sigma A_{\nu]}{}^\Gamma \,.
\end{equation}
Consequently, the variation of the Lagrangian is no longer a total
derivative when the gauge parameters are space-time dependent
functions. To obtain a variation that is equal to a total derivative,
one must include a new term in the Lagrangian \cite{deWit:1984px},
\begin{equation}
  \label{eq:cs-electric}
  \mathcal{L} =  \ft13  g\,
  \varepsilon^{\mu\nu\rho\sigma}\,X_{\Lambda\Sigma\Gamma}
  \,A_\mu{}^\Lambda A_\nu{}^\Sigma (\partial_\rho A_\sigma{}^\Gamma
  + \ft38 g\, X_{\Xi\Delta}{}^\Gamma \,A_\rho{}^\Xi A_\sigma{}^\Delta)
  \,.
\end{equation}
In the case of more general gauge group embeddings, this term is not
sufficient and extra terms will have to be introduced as soon as also
the magnetic gauge fields are present.

We now consider more general gauge groups without restricting
ourselves to electric charges. Therefore we must include both electric
gauge fields $A_\mu{}^\Lambda$ and magnetic gauge fields
$A_{\mu\,\Lambda}$.  Only a subset of these fields is usually involved
in the gauging, but the additional magnetic gauge fields could
conceivably lead to new propagating degrees of freedom. We will
see in due course how this is avoided.
The charges $X_{MN}{}^P$ correspond to a more general subgroup of the
duality group. Hence they must take values in the Lie algebra
associated with $\mathrm{Sp}(2n,\mathbb{R})$, which implies,
\begin{equation}
  \label{eq:sp-constraint}
  X_{M[N}{}^Q\,\Omega_{P]Q} =0\,.
\end{equation}
Furthermore we impose the representation constraint that was discussed
earlier. In this case the constraint implies that we
suppress a representation of the rigid symmetry group in $X_{MN}{}^P$
\cite{deWit:2005ub},
\begin{equation}
  \label{eq:lin}
  X_{(MN}{}^{Q}\,\Omega_{P)Q} =0
\Longrightarrow  \left\{
\begin{array}{l}
  X^{(\Lambda\Sigma\Gamma)}=0\,,\\[.2ex]
  2X^{(\Gamma\Lambda)}{}_{\Sigma}=
  X_{\Sigma}{}^{\Lambda\Gamma}\,, \\[.2ex]
  X_{(\Lambda\Sigma\Gamma)}=0\,,\\[.2ex]
  X_{(\Gamma\Lambda)}{}^{\Sigma}=
  X^{\Sigma}{}_{\Lambda\Gamma}\,.
\end{array}
\right.
\end{equation}
Observe that the generators $X_{\Lambda\Sigma}{}^\Gamma$ are not
necessarily
antisymmetric in $\Lambda$ and $\Sigma$; their antisymmetric part
appears in the field strengths \eqref{eq:nonabelian-FS}. The
symmetric part defines the tensor $Z^P{}_{MN}$ according to
\eqref{eq:def-Z}. In the case at hand we derive, using \eqref{eq:lin},
\begin{equation}
  \label{eq:Z-adjoint}
  Z^P{}_{MN} =  \tfrac12 \Omega^{PR}
  \Theta_R{}^\alpha \, t_{\alpha M}{}^Q \,\Omega_{NQ}\,,
\end{equation}
which shows that the symmetric index pair $(MN)$ of $Z^P{}_{MN}$ is
restricted to the adjoint representation of the rigid symmetry group
$\mathrm{G}$. Henceforth we use the notation,
\begin{equation}
  \label{eq:ZZ}
  X_{(MN)}{}^P = Z^P{}_{MN} = Z^{P,\alpha}\, d_{\alpha MN}\,,
\end{equation}
where
\begin{eqnarray}
  \label{eq:def-Z-d}
  d_{\alpha\, MN} &\equiv& (t_{\alpha})_M{}^P\, \Omega_{NP}\,,\nonumber\\
Z^{M,\alpha}&\equiv&\ft12\Omega^{MN}\Theta_{N}{}^{\alpha}
\quad
\Longrightarrow
\quad
\left\{
\begin{array}{rcr}
Z^{\Lambda{\alpha}} &=& \ft12\Theta^{\Lambda\alpha} \,,\\[1ex]
Z_{\Lambda}{}^{{\alpha}} &=& -\ft12\Theta_{\Lambda}{}^{\alpha} \,.
\end{array}
\right.
\end{eqnarray}
Note, however, that when the symmetry group $\mathrm{G}$ is not
simple, then the indices $\alpha,\beta,\ldots$ above will be
restricted to the invariant subgroup that acts on the vector fields.
The tensor $Z^{M,\alpha}$ takes only non-zero values for those indices
$\alpha$. Hence the 2-forms transform in a restricted
sub-representation (namely the adjoint representation) of the
symmetric tensor product, as we stressed earlier.

Let us now consider the closure constraint \eqref{eq:constraint2},
which gave rise to the orthogonality relation between the embedding
tensor and the tensor $Z$. In the case at hand this implies,
\begin{equation}
  \label{eq:constraint-eq}
  Z^{M\,\alpha}\, \Theta_{M}{}^\beta \, d_{\alpha PQ} = \tfrac12
  \Omega^{MN}\,\Theta_{M}{}^{\beta}\Theta_{N}{}^{\alpha}\,d_{\alpha PQ}
  = \mathcal{Q}_{(PQ)}{}^\beta  \approx  0 \;.
\end{equation}
In case the gauge group is embedded into a simple group then
$\Omega^{MN}\,\Theta_{M}{}^{\alpha}\Theta_{N}{}^{\beta}= 2\,
\Theta_{\Lambda}{}^{[\alpha}\Theta^{\Lambda \beta]}\approx0$. When the
group $\mathrm{G}$ is not simple, then the indices
$\alpha,\beta,\ldots$ refer only to the invariant subgroup that acts
on the vector fields. This shows that the charges induced by the
gauging must be mutually local, meaning that there exists an
electric/magnetic duality transformation such that all the non-trivial
gauge charges become electric. In the remainder of the text we will
assume that we are dealing with a simple symmetry group $\mathrm{G}$,
both for convenience and because this reflects the situation
encountered in the maximal supergravity theories.

Let us continue to derive the $p$-form hierarchy for this specific
example. First of all, one replaces the electric field strengths
$F_{\mu\nu}{}^\Lambda= 2\partial_{[\mu}A_{\nu]}{}^\Lambda$ in the
original ungauged Lagrangian by the electric components of the
modified field strengths \eqref{eq:modified-fs}, which in the case at
hand are written as,
\begin{equation}
  \label{eq:new-H-electric}
  \mathcal{H}_{\mu\nu}{}^M = \mathcal{F}_{\mu\nu}{}^M + g
  Z^{M,\alpha} B_{\mu\nu\,\alpha} \,.
\end{equation}
Here we used the definition
\begin{equation}
  \label{eq:B-alpha}
  B_{\mu\nu\alpha}=d_{\alpha MN}B_{\mu\nu}{}^{MN}\;.
\end{equation}
Furthermore one replaces the ordinary derivatives (on the matter
fields) by covariant ones, as specified in \eqref{eq:cov-derivative}.
Finally one adds a universal set of terms to the Lagrangian, which
generalize \eqref{eq:cs-electric}.  The Lagrangian thus takes the
following form \cite{deWit:2005ub},
\begin{equation}
  \label{eq:L+Ltop}
  \mathcal{L}_\mathrm{total} = \mathcal{L}_0 + \mathcal{L}_\mathrm{top}\,,
\end{equation}
where $\mathcal{L}_0$ is the original (ungauged) Lagrangian with the field
strengths $F_{\mu\nu}{}^\Lambda$ replaced by covariant field strengths
$\mathcal{H}_{\mu\nu}{}^\Lambda$, and the space-time derivatives
$\partial_\mu$ by covariant derivatives $D_\mu$. The term
$\mathcal{L}_\mathrm{top}$ is the generalization of
\eqref{eq:cs-electric}, and reads,
\begin{eqnarray}
  \label{eq:top-lagr}
  {\cal L}_\mathrm{top}&=&
  \tfrac{1}{8}g\varepsilon^{\mu\nu\rho\sigma}\Theta^{\Lambda
    \alpha}B_{\mu\nu
    \alpha}\left(2\partial_{\rho}A_{\sigma\Lambda}+
    gX_{MN\Lambda}A_{\rho}{}^MA_{\sigma}{}^N
    -\tfrac{1}{4}g\Theta_{\Lambda}{}^\beta
    B_{\rho \sigma\beta}\right)\nonumber\\
  &&{}
  +\tfrac{1}{3}g
  \varepsilon^{\mu\nu\rho\sigma}X_{MN\Lambda}A_{\mu}{}^MA_{\nu}{}^N
  \left(\partial_{\rho}A_{\sigma}{}^{\Lambda}+
    \tfrac{1}{4}gX_{PQ}{}^{\Lambda}A_{\rho}{}^PA_{\sigma}{}^Q\right)
    \nonumber\\
  &&{}
  +\tfrac{1}{6}g\varepsilon^{\mu\nu\rho\sigma}X_{MN}{}^{\Lambda}
  A_{\mu}{}^MA_{\nu}{}^N
  \left(\partial_{\rho}A_{\sigma\Lambda}
    +\tfrac{1}{4}gX_{PQ\Lambda}A_{\rho}{}^PA_{\sigma}{}^Q\right)\,.
\end{eqnarray}
The combined Lagrangian \eqref{eq:L+Ltop} is gauge invariant provided
the embedding tensor $\Theta_M{}^\alpha$ is constant and satisfies the
closure constraint \eqref{eq:constraint2}. The gauge transformations
for the 1- and 2-form gauge fields have already been defined earlier
in the context of an abstract $p$-form hierarchy, but as invariance
of the Lagrangian they acquire a different form
\cite{deWit:2005ub,deVroome:2007zd},
\begin{eqnarray}
  \label{eq:def-delta-A-B}
  \delta A_\mu{}^M &=& D_\mu\Lambda^M -g\,
  Z^{M,\alpha}\,\Xi_{\mu\,\alpha} \,, \nonumber\\
  \delta B_{\mu\nu\alpha} &=& 2\, D_{[\mu} \Xi_{\nu]\alpha}+2\,d_{\alpha
    MN}A_{[\mu}{}^M\delta A_{\nu]}{}^N- 2\,d_{\alpha
  MN}\,\mathcal{G}_{\mu\nu}{}^M\Lambda^N \nonumber\\
  &&{}
  - g\, Y_{\alpha,M}{}^{\beta}\Phi_{\mu\nu}{}^M{}_\beta\; ,
\end{eqnarray}
where
\begin{equation}
  \label{eq:G-gauged}
    {\cal G}_{\mu\nu\,\Lambda} = \varepsilon_{\mu\nu\rho\sigma}\,
  \frac{\partial \mathcal{L}_0}{\partial{{\cal H}}_{\rho\sigma}{}^\Lambda}
  \;,
\end{equation}
is the covariant version of \eqref{eq:def-G} and $\Xi_{\mu\alpha}=
d_{\alpha MN}\, \Xi_\mu{}^{MN}$. The covariant derivative of the
transformation parameter $\Xi_{\mu\,\alpha}$ equals $D_\mu
\Xi_{\nu\,{\alpha}}= \partial_\mu \Xi_{\nu\,\alpha} - g A_\mu{}^M
X_{M\alpha}{}^\beta \,\Xi_{\nu\,\beta}$ with $X_{M \alpha}{}^\beta=
-\Theta_M{}^\gamma f_{\gamma\alpha}{}^\beta$ the gauge group generator
in the adjoint representation of $\mathrm{G}$.  Observe that we have
also included a 3-form gauge transformations with parameter
$\Phi_{\mu\nu}{}^M{}_\beta$ in \eqref{eq:def-delta-A-B}. As long as
the closure constraint is satisfied, this transformation is
irrelevant, since the 2-form field appears in the Lagrangian
multiplied with $Z^{M,\alpha}$, which vanishes upon contraction with
$Y_{\alpha,M}{}^\beta$ by virtue of the closure constraint (c.f.
\eqref{eq:YY-aprox-Q}). This is the reason why the $p$-form hierarchy
is truncated at $p=2$.

For what follows, it is convenient to present alternative expressions
for the intertwining tensors \cite{deWit:2008ta,deWit:2008gc},
\begin{eqnarray}
  \label{eq:interpolater-consistency}
  Y_{\alpha,M}{}^\beta &=& t_{\alpha M}{}^N\,\Theta_N{}^\beta -
  X_{M}{}^\beta{}_\alpha \;,
  \nonumber\\[1.2ex]
  Y^M{}_{\alpha, PQ}{}^\beta  &=&{}- \delta_P{}^M\,Y_{\alpha,Q}{}^\beta
  -(X_P)_Q{}^{\beta,M}{}_\alpha \;,
  \nonumber\\[1.2ex]
  Y^{MN}{}_{\alpha, PQR}{}^\beta  &=&{}
  -\delta_P{}^M\,Y^N{}_{\alpha, QR}{}^\beta -
  (X_P)_{QR}{}^{\beta,MN}{}_\alpha    \;,
\end{eqnarray}
where the last terms $(X_M)$ denote the generators in the
representation conjugate to the representations associated with the
$p=2,3,4$ form fields. The above expressions are useful when
performing explicit calculations. Note that all intertwining tensors
are linear in the embedding tensor as well as in the generators
$(t_\alpha)_M{}^N$ or in the structure constants
$f_{\alpha\beta}{}^\gamma$. As was emphasized previously these tensors
do not cover all the (irreducible) representations that are allowed
by their index structure. For instance, the fact that the
representation constraint \eqref{eq:lin} remains zero under the action
of the rigid symmetry group, i.e. $\delta_\alpha
(X_{(MN}{}^{Q}\,\Omega_{P)Q}) =0$, implies that the following
contraction must vanish, $Y_{\alpha,(M}{}^\beta\,d_{\beta NP)}=0$.
Therefore the corresponding representations of the 3-form field
proportional to $\delta^M{}_{(N} d_{\alpha PQ)}$ times a symmetric
three-rank tensor will decouple from the hierarchy
\cite{deWit:2005hv,deWit:2008ta}.

It is possible to go beyond the $p=2$ truncation and introduce a 3-
and a 4-form field by making use of the observations at the end of
subsection \ref{sec:more-about-hierarchy}. Hence we introduce a 3-form
field $C_{\mu\nu\rho}{}^M{}_\alpha$ and a 4-form field
$D_{\mu\nu\rho\sigma}{}^{MN}{}_\alpha$. At the same time we relax the
constraints on the embedding tensor, which we allow to be a space-time
dependent field that transforms in the representation allowed by the
representation constraint \eqref{eq:lin}, but which will not be
subject to the closure constraint \eqref{eq:constraint2}. The 3- and
4-form fields then play the role of Lagrange multipliers that impose
the constancy of the embedding tensor and the closure constraint. Now
the Lagrangian takes the form
\begin{equation}
  \label{eq:L+Ltop+LC}
  \mathcal{L}_\mathrm{total} = \mathcal{L}_0 +
  \mathcal{L}_\mathrm{top} + \mathcal{L}_\mathrm{C} \,,
\end{equation}
where the first two terms are as before and the third term coincides
with \eqref{eq:LC} applied to this particular case,
\begin{equation}
  \label{eq:Lagr-CD}
  {\cal L}_C = - \tfrac{1}{48}
  \varepsilon^{\mu\nu\rho\sigma}
  \left\{4g\, C_{\nu\rho\sigma}{}^M{}_{\alpha}\,D_{\mu}\Theta_M{}^{\alpha}
  +g^2D_{\mu\nu \rho \sigma}{}^{MN}{}_{\alpha}
  \,\mathcal{Q}_{MN}{}^{\alpha}\right\}\, .
\end{equation}
Since the first two terms in \eqref{eq:L+Ltop+LC} are only gauge
invariant for a constant embedding tensor satisfying the closure
constraint, there will be new variations proportional to
$D_\mu\Theta_M{}^\alpha$ or $\mathcal{Q}_{MN}{}^\alpha$, which must be
absorbed by the variations of \eqref{eq:Lagr-CD}. This requirement
fixes the gauge transformation laws of the 3- and 4-form gauge fields.
Note that the sub-representation in the 3-form field proportional to
$\delta^M{}_{(N} d_{\alpha PQ)}$ decouples from the Lagrangian, in
view of the identity $D_{\mu}\Theta_{(M}{}^{\alpha}d_{\alpha NP)}=0$.
This is in accord with our discussion below 
\eqref{eq:interpolater-consistency}.

The calculation of these variations is tedious but straightforward. A
brief perusal of the variations shows that in the variation of
$\mathcal{L}_0$, these terms originate from new variations of the covariant
field strengths and the covariant derivatives. To see this
we first note that the formal closure of the gauge
algebra is affected,
\begin{equation}
  \label{eq:algebra-nonclosed}
  {[X_M, X_N]} = -X_{MN}{}^P X_P + \mathcal{Q}_{MN}{}^\alpha
  t_\alpha\,.
\end{equation}
Furthermore the Ricci identity \eqref{eq:ricci} is modified,
\begin{eqnarray}
  \label{eq:modified-Ricci}
  {[D_{\mu}, D_{\nu}]} &=& -g {\cal F}_{\mu \nu}{}^M X_M + \left[ 2 g
  A_{[\mu}{}^M D_{\nu]} \Theta_M{}^\alpha - g^2
  A_{[\mu}{}^M A_{\nu]}{}^N \mathcal{Q}_{MN}{}^\alpha\right]
  t_\alpha \nonumber\\
  &=& -g \mathcal{H}_{\mu \nu}{}^M X_M \nonumber\\
  &&{}
  + \left[ 2 g  A_{[\mu}{}^M D_{\nu]} \Theta_M{}^\alpha -
  g^2(A_{[\mu}{}^M A_{\nu]}{}^N -B_{\mu\nu}{}^{MN})
  \mathcal{Q}_{MN}{}^\alpha \right] t_\alpha\,.
\end{eqnarray}
The transformation of the field strengths $\mathcal{H}_{\mu\nu}{}^M$
will therefore become more complicated. Using \eqref{eq:def-delta-A-B}
one finds the following result,
\begin{equation}
  \label{eq:Htrans}
  \delta {\cal H}_{\mu\nu}{}^M = -g \Lambda^N X_{NP}{}^M
    \mathcal{G}_{\mu \nu}{}^P- g\Lambda^N X_{PN}{}^M
    (\mathcal{G}-\mathcal{H})_{\mu\nu}{}^P  +
    \Delta\mathcal{H}_{\mu\nu}{}^M\,,
\end{equation}
where $\Delta\mathcal{H}_{\mu\nu}{}^M$ contains the new variations
proportional to $D_\mu\Theta_M{}^\alpha$ or
$\mathcal{Q}_{MN}{}^\alpha$. These take the form,
\begin{eqnarray}
  \label{eq:Delta-H}
  \Delta \mathcal{H}_{\mu\nu}{}^M &=& -2g A_{[\mu}{}^N D_{\nu]}
  \Theta_N{}^\alpha \,t_{\alpha P}{}^M \Lambda^P +g\,\Xi_{[\mu \alpha}
  D_{\nu]}\Theta_N{}^\alpha\,\Omega^{MN} \nonumber\\
  &&{}
  +g^2A_{[\mu}{}^N A_{\nu]}{}^P
  \Lambda^Q\left(\mathcal{Q}_{NP}{}^\alpha t_{\alpha Q}{}^M +
  \mathcal{Q}_{QN}{}^\alpha
  t_{\alpha P}{}^M\right)\nonumber\\
   &&{}
   -\tfrac{1}{2}g^2 \Lambda^P \mathcal{Q}_{NP}{}^\alpha \Omega^{MN} B_{\mu
     \nu \alpha} -\tfrac{1}{2} g^2\Phi_{\mu \nu }{}^N{}_\alpha
   \Omega^{MP} \mathcal{Q}_{PN}{}^\alpha\,.
\end{eqnarray}
All these extra terms arise from the fact that the closure constraint
no longer holds, and that the embedding tensor and related quantities
are not constant and not gauge invariant anymore.

A similar result exists for the variation of a covariant derivative
on a field transforming according to some representation of the gauge
group,
\begin{equation}
  \label{eq:vari-covariant-der}
  \delta (D_\mu \Phi)  =  g\,\Lambda^M X_M D_\mu \Phi + \Delta(D_\mu)
  \Phi\,,
\end{equation}
where the second term is again proportional to
$D_\mu\Theta_M{}^\alpha$ or $\mathcal{Q}_{MN}{}^\alpha$. This term
takes the form,
\begin{equation}
  \label{eq:Delta-D}
  \Delta(D_{\mu}) =
  \left(g\Lambda^M\, D_{\mu}\Theta_M{}^{\alpha}-
   g^2\,\Lambda^M A_{\mu}{}^N\mathcal{Q}_{MN}{}^{\alpha}+
  g^2\,\Xi_{\mu}{}^{MN}\,\mathcal{Q}_{(MN)}{}^{\alpha}\right)t_{\alpha}
  \,.
\end{equation}

The effect of the new variations of the field strengths and covariant
derivatives thus lead to a new variation of the Lagrangian
$\mathcal{L}_0$,
\begin{eqnarray}
  \label{eq:extra-L0-var}
  \Delta {\cal L}_0 = - \tfrac{1}{4}\varepsilon^{\mu \nu \rho \sigma} {\cal
    G}_{\rho \sigma \Lambda} \, \Delta {\cal H}_{\mu\nu}{}^{\Lambda}
  +\frac{\delta {\cal L}_0}{\delta(D_{\mu}\Phi)}\, \Delta D_{\mu}
    \Phi\,.
\end{eqnarray}

What remains is to also evaluate the extra variations of the
Lagrangian $\mathcal{L}_\mathrm{top}$ defined in
\eqref{eq:top-lagr}, which are also proportional to
$D_\mu\Theta_M{}^\alpha$ or $\mathcal{Q}_{MN}{}^\alpha$. All these
variations must then cancel against the variations of
\eqref{eq:Lagr-CD}, provided one assigns the following gauge
transformations to the 3- and 4-form gauge fields (for conciseness, we
suppress the contributions coming from the second term in
\eqref{eq:extra-L0-var}), 
\begin{eqnarray}
  \label{eq:transf-C-D}
  \delta C_{\mu \nu \rho}{}^M{}_{\alpha} &=&
  3 \,D_{[\mu} \Phi_{\nu\rho]}{}^M{}_{\alpha}
  -3\, A_{[\mu}{}^M\delta B_{\nu \rho]\alpha}
  +3\,{\cal G}_{[\mu \nu}{}^M \Xi_{\rho]\alpha}
  +2\,d_{{\alpha} PQ}
  A_{[\mu}{}^M A_{\nu}{}^P \delta A_{\rho]}{}^Q  \nonumber\\
  &&{}
  -g \,Y^M{}_{{\alpha} ,PQ}{}^{\beta}\,
  \Upsilon_{\mu\nu \rho}{}^{PQ}{}_\beta\,,\nonumber \\[.2ex]
  %%%%
  \delta D_{\mu\nu\rho\sigma}{}^{MN}{}_{\alpha}&=&
  4\, D_{[\mu} \Upsilon_{\nu\rho\sigma]}{}^{MN}{}_\alpha
  + \Lambda^M {\cal H}_{\mu\nu\rho\sigma}{}^N{}_{\alpha}
  +3 (B_{[\mu\nu}{}^{(MN)} -2\,A_{[\mu}{}^M A_{\nu}{}^N) \,\delta
  B_{\rho\sigma]\alpha}
  \nonumber\\
  &&{}
  + 6\,\mathcal{G}_{[\mu \nu}{}^M \Phi_{\rho\sigma]}{}^N{}_{\alpha}
  +2\, d_{{\alpha}PQ} A_{[\mu}{}^M A_{\nu}{}^N A_{\rho}{}^P
  \delta A_{\sigma]}{}^Q +4\delta
  A_{[\mu}{}^MC_{\nu\rho\sigma]}{}^N{}_{\alpha}  \nonumber\\
  &&{}
  -g \,Y^{MN}{}_{{\alpha},PQR}{}^{\beta}\,
  \Pi_{\mu\nu\rho\sigma}{}^{PQR}{}_{\beta}\,.
\end{eqnarray}
Here, the transformations parameterized by the functions
$\Upsilon_{\mu\nu\rho}{}^{MN}{}_\alpha$ and
$\Pi_{\mu\nu\rho\sigma}{}^{MNP}{}_\alpha$ are associated with the
tensor gauge transformations of the 4- and 5-form fields. Of course,
5-form fields do not exist in a four-dimensional space-time, but this
transformation still has some effect as it acts by a shift
transformation on the 4-form field. The invariance of the Lagrangian
under this transformation is ensured by the identity
\eqref{eq:Y-Q-vanish}.

Furthermore, $\mathcal{H}_{\mu\nu\rho\sigma}{}^M{}_\alpha$ is the
covariant field strength associated with the 3-form field,
\begin{eqnarray}
  \label{eq:H-4}
  {\cal H}_{\mu\nu\rho\sigma}{}^M{}_{\alpha}& =&
  4\, D_{[\mu}C_{\nu\rho \sigma]}{}^M{}_{\alpha}+ 12\,
  d_{\alpha PQ} \,A_{[\mu}{}^MA_{\nu}{}^P
  \left(\mathcal{G}_{\rho\sigma]}{}^Q -\tfrac23
  \partial_{\rho}A_{\sigma]}{}^Q
  -\tfrac12 gX_{RS}{}^QA_{\rho}{}^RA_{\sigma]}{}^S\right)\nonumber\\
  &&{}
  -3\,B_{[\mu\nu\alpha}\left(2(\mathcal{G}- \mathcal{H})_{\rho\sigma]}{}^M
   + gZ^{M,\beta}B_{\rho\sigma]\beta}\right)+
 4\,g Y_{\alpha,P}{}^{\beta} A_{[\mu}{}^M \,C_{\nu\rho\sigma]}{}^P
  {}_{\beta}      \nonumber\\
  &&{}
  +g\,Y^M{}_{\alpha,PQ}{}^{\beta}\,
  D_{\mu\nu\rho\sigma}{}^{PQ}{}_{\beta}\,.
\end{eqnarray}
In defining this field strength we made use of our earlier
observation that certain sub-representations of the 3-form field
decouple from the theory.

At this point we have established the invariance of the Lagrangian
\eqref{eq:L+Ltop+LC}. Rather than giving further calculational details
which will be published elsewhere, we close with a number of comments.
First of all, we have already observed that the $p$-form
transformations obtained from the invariance of a certain Lagrangian,
differ from the transformations that are obtained along the lines
presented in subsection \ref{sec:gauge-deformations}. Nevertheless
there exists a relation between these two sets of transformation
rules. Namely they tend to be identical up to (Hodge) duality
relations between $p$-forms, some of which are satisfied as a result
of the field equations. However, as we know from
\cite{deWit:2008ta}, this relationship is only partially realized and
there exist some unexpected invariances in Lagrangians such as
\eqref{eq:L+Ltop+LC} that are necessary for obtaining a consistent
interpretation. This can be seen, in principle, by evaluating the commutator
algebra based on the theory above, which will close up to these
additional transformations.

Another intriguing feature of our result, which was noticed also in
\cite{deWit:2008ta}, is that for Lagrangians quadratic in derivatives,
one can, in principle, integrate out the embedding tensor field
$\Theta_M{}^\alpha$. Although the resulting Lagrangian tends to be
complicated and non-polynomial, it should encode all possible gaugings
of this type. Whether or not this intriguing observation has any
practical importance remains to be seen. We hope to return to these
and related questions in the future.

\vskip .4cm
\noindent
{\bf Note added in proof}: We include two recent papers on the
$p$-form hierarchy in four space-time dimensions. The first one
relates a modification of the representation constraint \eqref{eq:lin}
to anomaly cancellation \cite{DeRydt:2008hw}. The second one considers
the extension of the hierarchy with 3- and 4-form fields, and is
directly related to the material presented in section
\ref{sec:hierarchyin4d} \cite{Bergshoeff:2009ph}.

%%%%%%%%%%%%%%%%%%%%%%%%%%%%%%%%%%%%%%%%%%%%%%%%%%%%%%%%%%%%%
%\newpage
\vspace{8mm}
%%%%%%%%%%%%%%%%%%%%%%%%%%%%%%%%%%%%%%%%%%%%%%%%%%%%%%%%%%%%%
\noindent
{\bf Acknowledgement}\\
\noindent
We are grateful to Hermann Nicolai and H. Samtleben for discussions.
The work of M.v.Z. is part of the research program of the `Stichting
voor Fundamenteel Onderzoek der Materie (FOM)', which is financially
supported by the `Nederlandse Organisatie voor Wetenschappelijk
Onderzoek (NWO)'. This work is also supported by NWO grant 047017015.
\bigskip

%%%%%%%%%%%%%%%%%%%%%%%%%%%%%%%%%%%%%%%%%%%%%%%%%%%%%%%%%%%%
%%%%%%%%%%%%%%%%%%%%%%%%%%%%%%%%%%%%%%%%%%%%%%%%%%%%%%%%%%%%
% ---- Bibliography ----
%%%%%%%%%%%%%%%%%%%%%%%%%%%%%%%%%%%%%%%%%%%%%%%%%%%%%%%%%%%

%%%%%%%%%%%%%%%%%%%%%%%%%%%%%%%%%%%%%%%%%%%%%%%%%%%%%%%%%%%%%%%%%%%
%%%%%%%%%%%%%%%%%%%%%%%%%%%%%%%%%%%%%%%%%%%%%%%%%%%%%%%%%%%%%%%%%%%
%
%%%%%%%%%%%%%%%%%%%%%%%%%%%%%%%%%%%%%%%%%%%%%%%%%%%%%

\begin{thebibliography}{99}
%
\bibitem{deWit:2004nw} B.~de~Wit, H.~Samtleben and M.~Trigiante,
     {\it The maximal $D=5$ supergravities}, Nucl.\ Phys.\ {\bf B716}
     (2005) 215, {\tt hep-th/0412173}.
     %%CITATION = NUPHA,B716,215;%%
%
  \bibitem{deWit:2005hv} B.~de~Wit and H.~Samtleben, {\it Gauged
      maximal supergravities and hierarchies of nonabelian
      vector-tensor systems}, { Fortsch. Phys.} {\bf 53} (2005) 442,
    {\tt hep-th/0501243}.
%%CITATION = FPYKA,53,442;%%
%
   \bibitem{deWit:2005ub} B.~de~Wit, H.~Samtleben and M.~Trigiante, {\it
       Magnetic charges in local field theory}, JHEP {\bf 09} (2005)
     016, {\tt hep-th/0507289}.
     %%CITATION = JHEPA,0509,016;%%
%
\bibitem{Nicolai:2000sc} H. Nicolai and H. Samtleben, {\it Maximal
    gauged supergravity in three dimensions}, Phys.\ Rev.\ Lett.\ {\bf
    86} (2001) 1686, {\tt hep-th/0010076};
  %%CITATION = PRLTA,86,1686;%%
%
\bibitem{Nicolai:2001sv} H.~Nicolai and H.~Samtleben,
  {\it Compact and noncompact gauged
    maximal supergravities in three-dimensions}, JHEP {\bf 0104} (2001)
  022, {\tt hep-th/0103032}.
  %%CITATION = JHEPA,0104,022;%%
%
\bibitem{deWit:2008ta}
B. de Wit, H. Nicolai and H. Samtleben, {\it Gauged
     supergravities, tensor hierarchies, and M-theory}, JHEP {\bf 02}
     (2008) 044, {\tt arXiv 0801.1294 [hep-th]}.
   %%CITATION = 0801.1294;%%
%
\bibitem{Cremmer:1978km} E.~Cremmer, B.~Julia and J.~Scherk, {\it
  Supergravity theory in 11 dimensions}, Phys.Lett. {76B} (1978) 409.
   %%CITATION = PHLTA,B76,409;%%
%
\bibitem{KK} Th. Kaluza, {\it Zum Unit\"atsproblem in der Physik},
  Sitzungsber.Preuss. Akad. Wiss. Berlin {\bf 1921}: 966,\\
  O. Klein, {\it Quantentheorie und f\"unfdimensionale
  Relativit\"atstheorie}, Z. f. Physik {\bf 37} (1926) 895.
%
\bibitem{deWit:2003ja} B. de Wit, I. Herger and H. Samtleben, {\it
    Gauged locally supersymmetric D = 3 nonlinear sigma models},
     Nucl. Phys. {\bf B671} (2003) 175, {\tt hep-th/0307006}.
    %%CITATION = HEP-TH/0307006;%%%
%
\bibitem{deWit:2004yr}
     B. de Wit, H. Nicolai and H. Samtleben, {\it Gauged supergravities
     in three dimensions: A panoramic overview}, Proc. 27th Johns
     Hopkins Workshop on Current Problems in Particle Theory:
     Symmetries and Mysteries of M-Theory, Goteborg, Sweden, 24-26 Aug
     2003, {\tt hep-th/0403014}.
    %%CITATION = HEP-TH/0403014;%%
%
\bibitem{Samtleben:2005bp} H.~Samtleben and M.~Weidner, {\it The
       maximal $D=7$ supergravities}, Nucl.\ Phys.\ {\bf B725} (2005)
     383, {\tt hep-th/0506237}.
     %%CITATION = NUPHA,B725,383;%%
%
\bibitem{deWit:2007mt} B. de Wit, H. Samtleben M. Trigiante, {\it The
     maximal D = 4 supergravities}, JHEP, {\bf 06} (2007) 049,
     {\tt arXiv:0705.2101 [hep-th]}
     %%CITATION = ARXIV:0705.2101;%%
%
\bibitem{Bergshoeff:2007ef} E.~Bergshoeff, H.~Samtleben and E. Sezgin,
  {\it The gaugings of maximal D=6 supergravity}, JHEP {\bf 03} (2008) 068,
       {\tt arXiv:0712.4277 [hep-th]}.
     %%CITATION = ARXIV:0712.4277;%%
%
\bibitem{deWit:2008gc} B. de Wit and H. Samtleben, {\it The end of the
     p-form hierarchy}, JHEP {\bf 08} (2008) 015, {\tt arXiv 0805.4767
     [hep-th]}.
     %%CITATION = 0805.4767;%%
%
\bibitem{Scherk:1979zr} J. Scherk and J.H. Schwarz, {\it How to Get
     Masses from Extra Dimensions}, Nucl. Phys. {\bf B153} (1979)
     61-88.
     %%CITATION = NUPHA,B153,61;%%
%
\bibitem{Obers:1998fb} N.A. Obers and B. Pioline, {\it U-duality and
     M-theory}, Phys. Rept. {\bf 318} (1999) 113-225", {\tt
     hep-th/9809039}.
     %%CITATION = HEP-TH/9809039;%%
%
\bibitem{deWit:2000wu} B. de Wit and H. Nicolai, {\it Hidden
     symmetries, central charges and all that},
     Class. Quant. Grav. {\bf 18} (2001) 3095-3112, {\tt hep-th/0011239}.
     %%CITATION = HEP-TH/0011239;%%
%
\bibitem{Elitzur:1997zn} S. Elitzur, A. Giveon, D. Kutasov and
                  E. Rabinovici, {\it Algebraic aspects of matrix
                  theory on $T^d$}, Nucl. Phys. {\bf B509} (1998)
                  122-144, {\tt hep-th/9707217}.
     %%CITATION = HEP-TH/9707217;%%
%
\bibitem{deWit:1988ig} B.~de Wit, J.~Hoppe and H.~Nicolai, {\it On
       the quantum mechanics of supermembranes}, Nucl. Phys. {\bf B305}
     (1988) 545.
     %%CITATION = NUPHA,B305,545;%%
%
\bibitem{Banks:1996vh} T.~Banks, W.~Fischler, S.H.~Shenker and
  L.~Susskind, {\it M-Theory as a matrix model: A conjecture}, Phys.
  Rev. {\bf D55} (1997) 5112, {\tt hep-th/9610043}.
  %%CITATION = HEP-TH/9610043;%%
%
\bibitem{Hull:1994ys} C.~Hull and P.K.~Townsend, {\it Unity of
    superstring dualities}, Nucl. Phys. {\bf B438} (1995) 109, {\tt
    hep-th/9410167}.
  %%CITATION = HEP-TH/9410167;%%
%
\bibitem{Iqbal:2001ye} A. Iqbal, A. Neitzke and C. Vafa, {\it A
     mysterious duality}, Adv. Theor. Math. Phys. {\bf 5} (2002)
     769-808, {\tt hep-th/0111068}.
     %%CITATION = HEP-TH/0111068;%%
%
 \bibitem{West:2004kb} P.C. West, {\it E(11) origin of brane
  charges and U-duality multiplets}, JHEP {\bf 08} (2004) 052, {\tt
  hep-th/0406150}.
     %%CITATION = HEP-TH/0406150;%%
%
\bibitem{Riccioni:2007au} F. Riccioni and P. West, {\it The E(11)
     origin of all maximal supergravities} JHEP {\bf 07} (2007) 063,
     {\tt arXiv:0705.0752 [hep-th]}.
     %%CITATION = ARXIV:0705.0752;%%
%
\bibitem{Bergshoeff:2007qi} E.A. Bergshoeff, I. De Baetselier and
                  T.A. Nutma, {\it E(11) and the embedding tensor},
                  JHEP {\bf 09} (2007) 047, {\tt arXiv:0705.1304
                  [hep-th]}.
     %%CITATION = ARXIV:0705.1304;%%
%
\bibitem{Riccioni:2007ni} F. Riccioni and P. West, {\it E(11)-extended
     spacetime and gauged supergravities}, JHEP {\bf 0802}, 039
     (2008), {\tt arXiv:0712.1795 [hep-th]}.
     %%CITATION = ARXIV:0712.1795;%%
%
\bibitem{Bergshoeff:2008qd}
  E.A.~Bergshoeff, O.~Hohm and T.A.~Nutma,
  {\it A note on ${\rm E}_{11}$ and three-dimensional gauged supergravity},
  JHEP {\bf 0805} (2008) 081,
  {\tt arXiv:0803.2989 [hep-th]}.
  %%CITATION = ARXIV:0803.2989;%%
%
\bibitem{Englert:2007qb} F.~Englert, L.~Houart, A.~Kleinschmidt,
  H.~Nicolai and T.~Nassiba, {\it An $E_9$ multiplet of BPS states},
  JHEP {\bf 05} (2007) 065, {\tt hep-th/0703285}.
  %%CITATION = HEP-TH/0703285;%%
%
\bibitem{Bergshoeff:2008xv} E.A. Bergshoeff, O. Hohm, A. Kleinschmidt,
  H. Nicolai, T.A. Nutma, and J. Palmkvist, {\it $\mathrm{E}_{10}$ and
    gauged maximal supergravity}, JHEP {\bf 01} (2009) 020, {\tt
    arXiv:0810.5767 [hep-th]}.
     %%CITATION = 0810.5767;%%%
%
\bibitem{Samtleben:2007an} H.~Samtleben and M.~Weidner, {\it Gauging
    hidden symmetries in two dimensions}, JHEP {\bf 08} (2007) 076,
    {\tt arXiv:0705.2606 [hep-th]}.
    %%CITATION = ARXIV:0705.2606;%%
%
\bibitem{Schon:2006kz} J.~Sch\"on and M.~Weidner, {\it Gauged $N = 4$
    supergravities}, JHEP {\bf 05} (2006) 034, {\tt hep-th/0602024}.
  %%CITATION = JHEPA,0605,034;%%
%
\bibitem{Derendinger:2007xp} J.P.~Derendinger,
                P.M.~Petropoulos and N.~Prezas, {\it Axionic symmetry
                  gaugings in \mbox{N = 4} supergravities and their
                  higher-dimensional origin}, Nucl.\ Phys.\ {\bf B785}
                (2007) 115, {\tt arXiv:0705.0008 [hep-th]}.
 %%CITATION = NUPHA,B785,115;%%
%
\bibitem{deVroome:2007zd} M.~de Vroome and B.~de Wit, {\it Lagrangians
    with electric and magnetic charges in $N=2$ supersymmetric gauge
    theories}, JHEP {\bf 08} (2007) 064, {\tt arXiv:0707.2717
    [hep-th]}.
    %%CITATION = ARXIV:0707.2717;%%
%
\bibitem{deWit:2001pz}
  B.~de~Wit, {\it Electric-magnetic duality in supergravity},
  {Nucl. Phys. Proc. Suppl.} {\bf 101} (2001) 154,
  {{\tt hep-th/0103086}}.
  %%CITATION = HEP-TH 0103086;%%.
%
\bibitem{deWit:1984px} B.~de~Wit, P.~G. Lauwers, and A.~Van~Proeyen,
  \emph{ Lagrangians of {N}=2 supergravity-matter systems}, { Nucl.
    Phys.} {\bf B255} (1985) 569.
%%CITATION = NUPHA,B255,569;%%.
%
\bibitem{DeRydt:2008hw}
    J. De Rydt, T.T.  Schmidt, M. Trigiante, A. Van Proeyen and
    M. Zagermann, {\it Electric/magnetic duality for chiral gauge
     theories with anomaly cancellation}, JHEP {\bf 12} (2008) 105",
    {\tt arXiv:0808.2130 [hep-th]}.
     %%CITATION = 0808.2130;%%
%
\bibitem{Bergshoeff:2009ph}
    E.A. Bergshoeff, J. Hartong, O. Hohm, M. Huebscher and T. Ortin,
    {\it Gauge theories, duality relations and the tensor hierarchy},
    {\tt arXiv:0901.2054 [hep-th]}.
     %%CITATION = 0901.2054;%%
%
%%%%%%%%%%%%%%%%%%%%%%%%%%%%%%%%%%%%%%%%%%%%%%%%%%%%%%%%%%%%%%%
%
%
%%%%%%%%%%%%%%%%%%%%%%%%%%%%%%%%%%%%%%%%%%%%%%%%%%%%%%%%%%%%%%%%%%%
\end{thebibliography}
\end{document}